\documentclass[%
twocolumn,
 amsmath,
 amssymb,
 aps,
 pra,
 reprint,
nofootinbib,
]{revtex4-2}

\usepackage[utf8]{inputenc}
\usepackage{datetime} %Only for printing time of compilation on drafts

\usepackage{dcolumn} % Align to decimal point in table

\usepackage{xcolor}
\usepackage[%
  colorlinks=true,
  urlcolor=blue,
  linkcolor=blue,
  citecolor=blue
]{hyperref}

\usepackage{graphicx}

\usepackage[caption=false,position=above]{subfig}
\usepackage{siunitx}
\usepackage{CJK}

\newcommand{\Erion}[1]{\textsuperscript{#1}Er\textsuperscript{3+}}

\begin{document}

\title{Microwave--optical double resonance in a erbium-doped whispering-gallery-mode resonator}

\author{Li Ma}
\author{Luke S. Trainor}
\author{Gavin G. G. King}
\author{Harald G. L. Schwefel}
\author{Jevon J. Longdell}\email{jevon.longdell@otago.ac.nz}
\affiliation{Dodd-Walls Centre for Photonic and Quantum Technologies, New Zealand}
\affiliation{Department of Physics, University of Otago, Dunedin, New Zealand}

\date{\today{}; \currenttime}

\begin{abstract}
    We showcase an erbium-doped whispering-gallery-mode resonator with optical modes that display intrinsic quality factors better than \num{e8} (linewidths less than $\SI{2}{\mega\hertz}$), and coupling strengths to collective erbium transitions of up to $2\pi\times\SI{1.2}{\giga\hertz}$ -- enough to reach the ensemble strong coupling regime.
    Our optical cavity sits inside a microwave resonator, allowing us to probe the spin transition which is tuned by an external magnetic field.
    We show a modified optically detected magnetic resonance measurement that measures
    population transfer by a change in coupling strength rather than absorption coefficient.
    This modification was enabled by the strong coupling to our modes, and allows us to optically probe the spin transition detuned by more than the inhomogeneous linewidth.
    We contrast this measurement with electron paramagnetic resonance to experimentally show that our optical modes are confined in a region of large microwave magnetic field and we explore how such a geometry could be used for coherent microwave--optical transduction.
\end{abstract}

\maketitle

\section{introduction}

Rare earth ions in solids at low temperatures provide a unique set of capabilities for coherently manipulating information. The $4f-4f$ transitions have very narrow homogeneous \cite{bottger_effects_2009,sun_recent_2002,equall_ultraslow_1994} and inhomogeneous \cite{ahlefeldt_ultranarrow_2016,macfarlane_inhomogeneous_1992} optical linewidths, as well as long lived coherence times for both nuclear \cite{zhong_optically_2015} and electron spin transitions \cite{ortu_simultaneous_2018,rakonjac_long_2020}. Large optical absorption is possible, despite the weak optical oscillator strengths, even to the point where negative refractive index could be possible \cite{berrington_negative_2022}. Furthermore, the use of erbium dopants gives access to an optical transition at \SI{1.5}{\micro\metre}, in the center of the low-loss wavelength region for optical fibers.

The properties of rare earth ion dopants have enabled a number of advances in quantum memories \cite{lago-rivera_telecom-heralded_2021, zhong_optically_2015}, and classical signal processing \cite{harris_multigigahertz_2006,berger_rf_2016}. A particular advantage of using dopants over free space atoms is increased flexibility when coupling the dopants to optical resonators. The coupling of rare earth ion dopants to nano-photonic resonators has enabled single site detection \cite{kolesov_optical_2012} as well as control and readout of single dopants \cite{kindem_control_2020}. Meanwhile, the coupling of ensembles to macroscopic resonators has been investigated to improve quantum memories \cite{jobez_cavity-enhanced_2014}, and microwave--optical transduction \cite{williamson_magneto-optic_2014,miyazono_coupling_2016,fernandez-gonzalvo_cavity-enhanced_2019}.

Whispering gallery mode (WGM) resonators combine extremely high quality ($Q$) factors and moderately small mode volumes in a convenient monolithic form factor \cite{strekalov_nonlinear_2016}. The guiding of light is provided by total internal reflection from the curved surface of the resonator's rim. Thus the resonators can operate over a wide range of wavelengths and can be made of any optical material that is low loss and for which smooth surfaces can be prepared. For example, recently $Q$-factors of \num{e9} have been demonstrated \cite{norman_measuring_2022} for yttrium orthosilicate (YSO) -- a popular host for cryogenic rare earth ion dopants because of its low density of spins.

The flexibility in material choice for WGM resonators has allowed numerous nonlinear optical processes like second-harmonic generation \cite{kuo_second-harmonic_2014,furst_second-harmonic_2015,trainor_selective_2018}, Kerr solitons \cite{herr_temporal_2014,webb_experimental_2016} and optical parametric oscillation \cite{meisenheimer_continuous-wave_2017} in wide frequency ranges.
The high $Q$-factors have enabled efficient demonstrations of photon pair sources \cite{fortsch_highly_2015} and allowed squeezing measurements of optical parametric oscillation  operated far above threshold \cite{furst_quantum_2011}.
Interfacing a microwave field with the resonators has allowed for electro-optic demonstrations such as efficient single-sideband modulation \cite{savchenkov_tunable_2009} and electro-optic frequency combs \cite{zhang_broadband_2019,rueda_resonant_2019}.
Coupling to ensemble spin states has previously been achieved through the interaction with magnon modes of yttrium iron garnet spheres \cite{haigh_selection_2018}.

Cryogenic rare earths in WGM resonators are relatively unexplored, in spite of the promises of very $Q$-factors for both the resonators and the ions.  It has been shown that being near to the surface in a WGM resonator doesn't adversely effect the ions spectral properties \cite{mcauslan_coherent_2011}.

Optical photons have much higher energies and couple to matter more strongly than radio-frequency or microwave photons. This makes optically detected magnetic resonance (ODMR) a powerful technique for probing solid state spins.
ODMR works by looking at spin-state dependent changes in the amount of optical absorption or fluorescence level, as either property depends on the population balance
\cite{carbonera_optically_2009,suter_optical_2020}.
ODMR is a more sensitive technique than EPR, not only due to the higher energy probe photons, but also because thermal noise is much lower at optical frequencies.
The optical detection significantly improves the sensitivity of magnetic resonance spectroscopy and increases the fidelity and spatial resolution of spin initialization and readout \cite{chernobrod_spin_2005,jacques_dynamic_2009,siyushev_coherent_2014}.

In our resonator we perform ODMR by monitoring changes in the materials dispersion rather than absorption.
This dispersive measurement is allowed by strong coupling between an optical resonator mode and an ensemble mode of the rare-earth ions.
Through the change of a polariton's frequency we measure the change in coupling due to exciting more ions into the upper spin state.
Large coupling strengths mean we can make this measurement when detuned by more than the inhomogeneous linewidth of the bare ions.
Thereby the amplitude of ensemble excitation by the optical frequency is reduced.

This work was motivated by the challenge of transferring quantum states from microwave to optical frequencies \cite{lambert_coherent_2020,lauk_perspectives_2020}. Superconducting qubits operate at frequencies of about \SI{10}{\giga\hertz} and hence must operate at temperatures of tens of millikelvins in order to not be swamped by thermal photons.
Interconnecting qubits in different cryostats is therefore difficult as a direct link would need to operate at a similar temperature \cite{magnard_microwave_2020}.
The suggested alternative is coherently transducing the microwave photons to optical frequencies where they can propagate through room temperature links with minimal thermal background.
By controlling the spin-state energy splitting with an external magnetic field, we can engineer a three-level system with appropriate energy levels for the transduction.
This modified ODMR helps us with system characterisation as the optical probe naturally measures only those ions within the optical mode volume.

\section{Experimental setup}

    \begin{figure}
        \subfloat[Microwave and optical apparatus]{\includegraphics[width=.9\columnwidth]{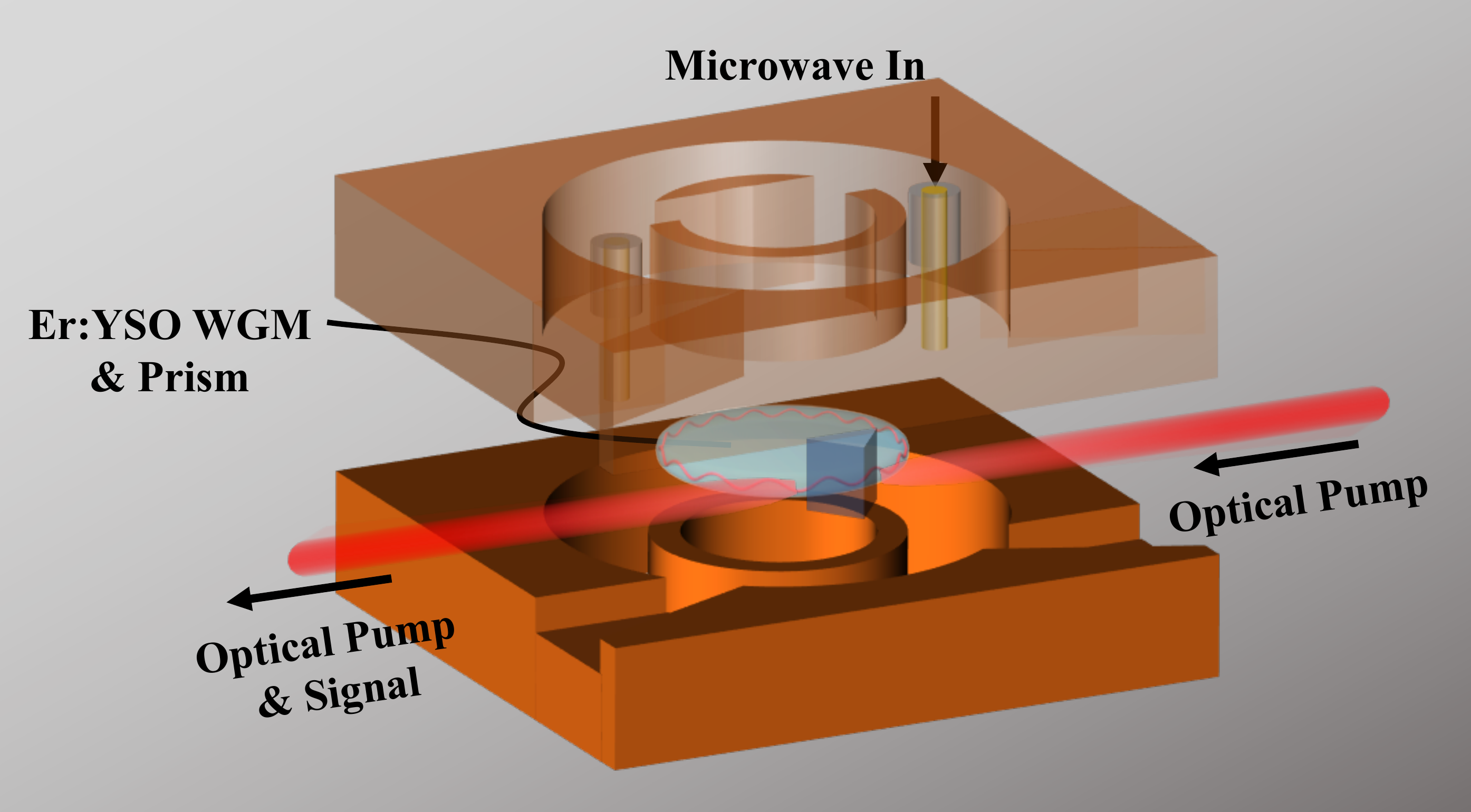}\label{fig:device:a}}\vspace{1pt}
        \\
        \subfloat[Energy-level diagram]{\includegraphics[width=.85\columnwidth]{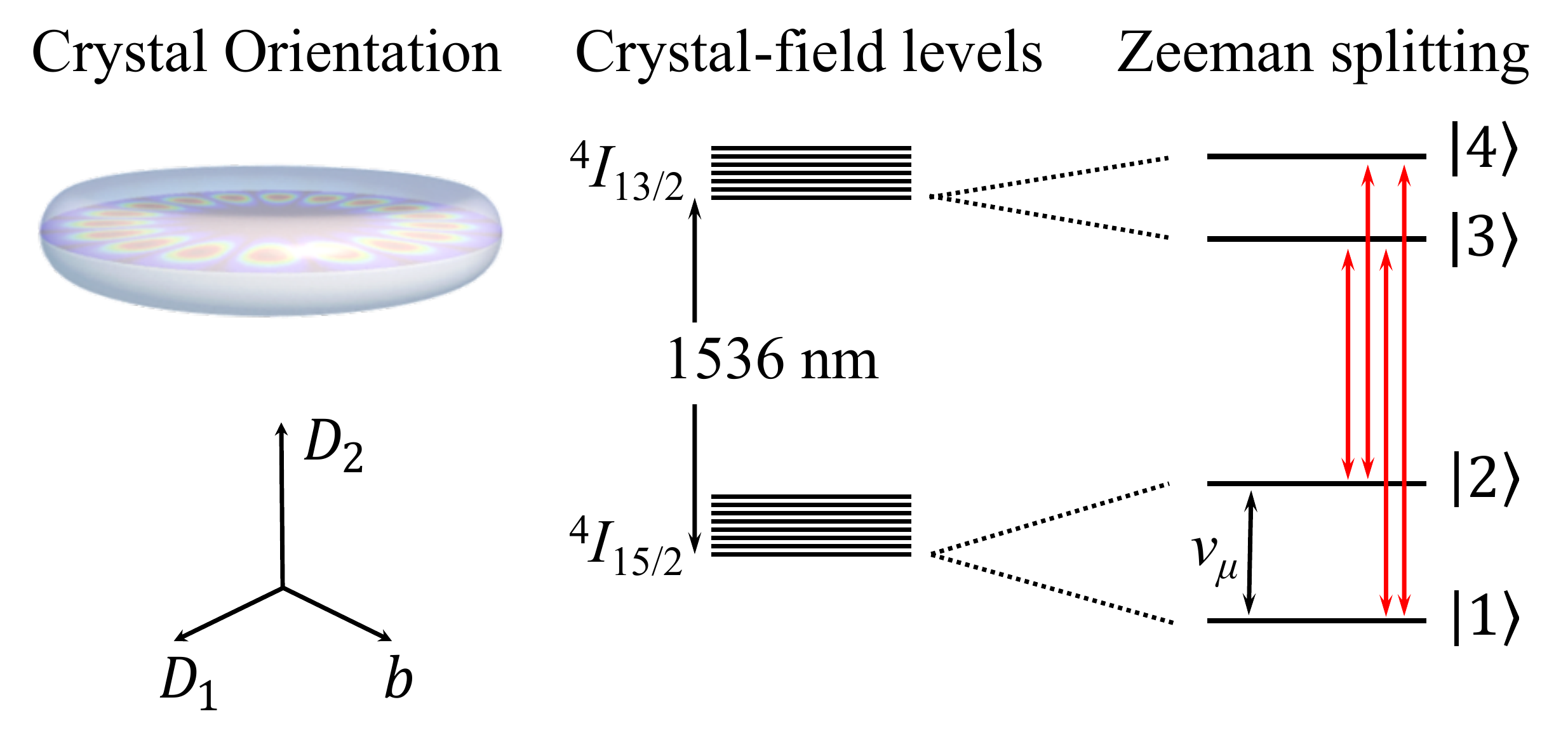}}\vspace{1pt}
        \caption{
        (a) A schematic of the ODMR apparatus. The microwave resonator (orange) is a shielded copper cavity with the top and bottom rings that clamp onto the WGM resonator.
        % A slot in the top ring is used for the phase matching between the microwave magnetic field and optical field.
        Two coaxial pins (gold) are used for coupling microwave in and out the cavity. The WGM resonator (light blue disc) is placed between the copper rings, with a prism (dark blue) that couples the optical pump (red) to the WGM resonator via evanescent coupling. (b) Energy-level diagram of the erbium ions. With an external magnetic field, the crystal levels of $^4I_{13/2}$ and $^4I_{15/2}$ have Zeeman split.
        $\nu_\mu$ shows the microwave transition that interacts with our microwave resonator; also shown are the four optical transitions that interact with the optical resonator in ascending order of energy.
        }
        \label{fig:device}
    \end{figure}
    
Our WGM resonator is made from YSO doped with isotopically enriched erbium-170 at \SI{50}{ppm}.
The resonator is a disk of major radius $R = \SI{2.1}{\milli\meter}$ and thickness of \SI{0.5}{\milli\meter}, with its curved sidewall surface finely polished by diamond slurry \cite{grudinin_ultrahigh_2006}. The crystal $D_2$ axis is normal to the disk plane.
It is well known that YSO has yttrium ions sitting in two different crystallographic sites each with two different orientations \cite{sun_magnetic_2008}. Here we address the erbium ions at site 1 with the optical transition around $\nu_0=\SI{195116.7}{\giga\hertz}$.

The optical pump couples into the WGM resonator by a right-isosceles gadolinium gallium garnet prism via evanescent coupling.
The crystal's $b$ axis points from the disk center to the coupling prism.
The prism is moved by a linear piezo stage to accurately tune the extrinsic optical coupling rate in situ. 
Such a prism conveniently allows for coupling with a straight-through geometry in the cryostat.
We label WGMs with their electric field parallel to the $D_2$ axis as the transverse electric (TE) polarization; those polarized in plane are transverse magnetic (TM).
TM modes have lower coupling contrast than TE modes due to their different local refractive indices at the coupling point.

The WGM resonator sits inside a microwave resonator---see Fig.~\ref{fig:device}(a)---which uses two metal rings, above and below the optical WGM resonator, to confine the microwave mode near the optical mode volume.  The resonator is based on previous designs \cite{rueda_efficient_2016}; one modification here is that to allow a greater tuning range, the whole floor of the resonator was movable rather than just a tuning pin. Another modification is that the top ring has an asymmetric slot, designed to break the near two-fold rotational symmetry of the microwave magnetic field and relax the phase matching requirements for future upconversion experiments.
A layer of indium is used between the metal rings and WGM resonator to reduce the gap between the two and better clamp the resonator.

The combined optical and microwave resonators were mounted in a home-built cryostat. The base temperature is \SI{4}{\kelvin}, which can be temporarily reduced to \SI{2.9}{\kelvin} by a helium evaporation system.
A 3D vector magnet applies a tunable external magnetic field along the $D_2$ axis.
The magnetic field causes Zeeman splitting of the ${}^4I_{13/2}$ and ${}^4I_{15/2}$ transition which we examine here, see Fig.~\ref{fig:device}(b).

\begin{figure}
    \includegraphics[width=1\columnwidth]{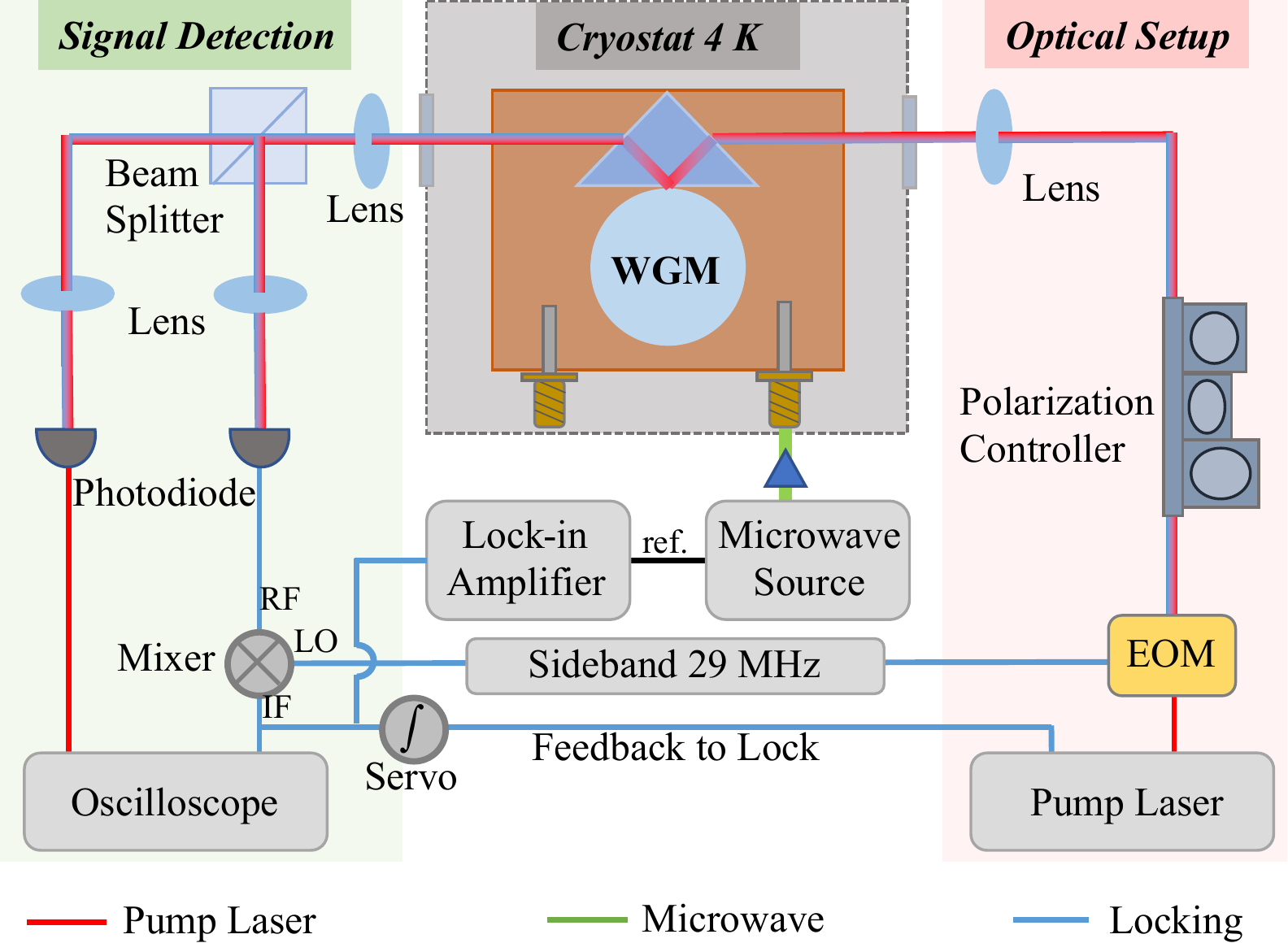}
    \caption{A schematic of the experimental setup and measurement. A tunable laser of \SI{1536}{\nano\meter} acts as an optical pump, with an EOM for relative frequency measurement and a polarization controller for WGM coupling.
    The microwave transition is driven by a microwave source with frequency around \SI{12.155}{\giga\hertz}. The reflected light is collected and detected by two photodiodes for optical monitoring and locking. When ODMR is being measured the microwave source is amplitude modulated and the resulting optical mode frequency oscillation is detected by a lock-in amplifier. The color lines in the bottom indicate the different frequencies of pump laser, microwave, and sideband modulation.}
    \label{fig:setup}
\end{figure}

A tunable fiber laser around \SI{1536}{\nano\metre} is focused by lenses outside the cryostat onto the prism--resonator interface for evanescent coupling, and the reflected light is collected on a photodiode.
An electro-optic phase modulator (EOM) and second photodiode are used for generation of a Pound--Drever--Hall (PDH) error signal.
The EOM also allows for accurate relative frequency calibration for measuring optical linewidths and free spectral ranges (FSRs).
At base temperature, we measured the intrinsic linewidth of a TE mode at \SI{195126.5\pm0.4}{\giga\hertz} ($\approx\SI{1536.4}{\nano\metre}$) -- about \SI{10}{\giga\hertz} away from the optical transition of \Erion{} -- to be \SI{1.32\pm0.10}{\mega\hertz}, corresponding to a quality factor of \num{1.48\pm0.12e8}. A TM mode about \SI{0.9}{\giga\hertz} lower in frequency had a quality factor of \num{1.07\pm0.07e8}. Meanwhile the optical modes have a free spectral range (FSR) of around \SI{12.3}{\giga\hertz}, and the tunable range of the microwave resonance frequency is approximately \SI{0.15}{\giga\hertz} centered around \SI{12.15}{\giga\hertz}. By moving the microwave tuning plate in and out, we can choose the microwave resonance frequency and make it match the optical FSR.

The coupling of optical modes to the erbium ions was measured by scanning the laser via its internal piezo  and tuning the external magnetic field to linearly change the erbium ions' optical transitions.
Figure 3(a) shows a 2D map of the optical spectrum with an applied magnetic field nearly along the resonator's $D_2$ axis. Each line is an optical mode with its corresponding anti-crossings where it approaches degeneracy with the erbium transitions.
The four straight dashed lines indicate the transitions between different energy levels.

\begin{figure}
    \includegraphics[width=1\columnwidth]{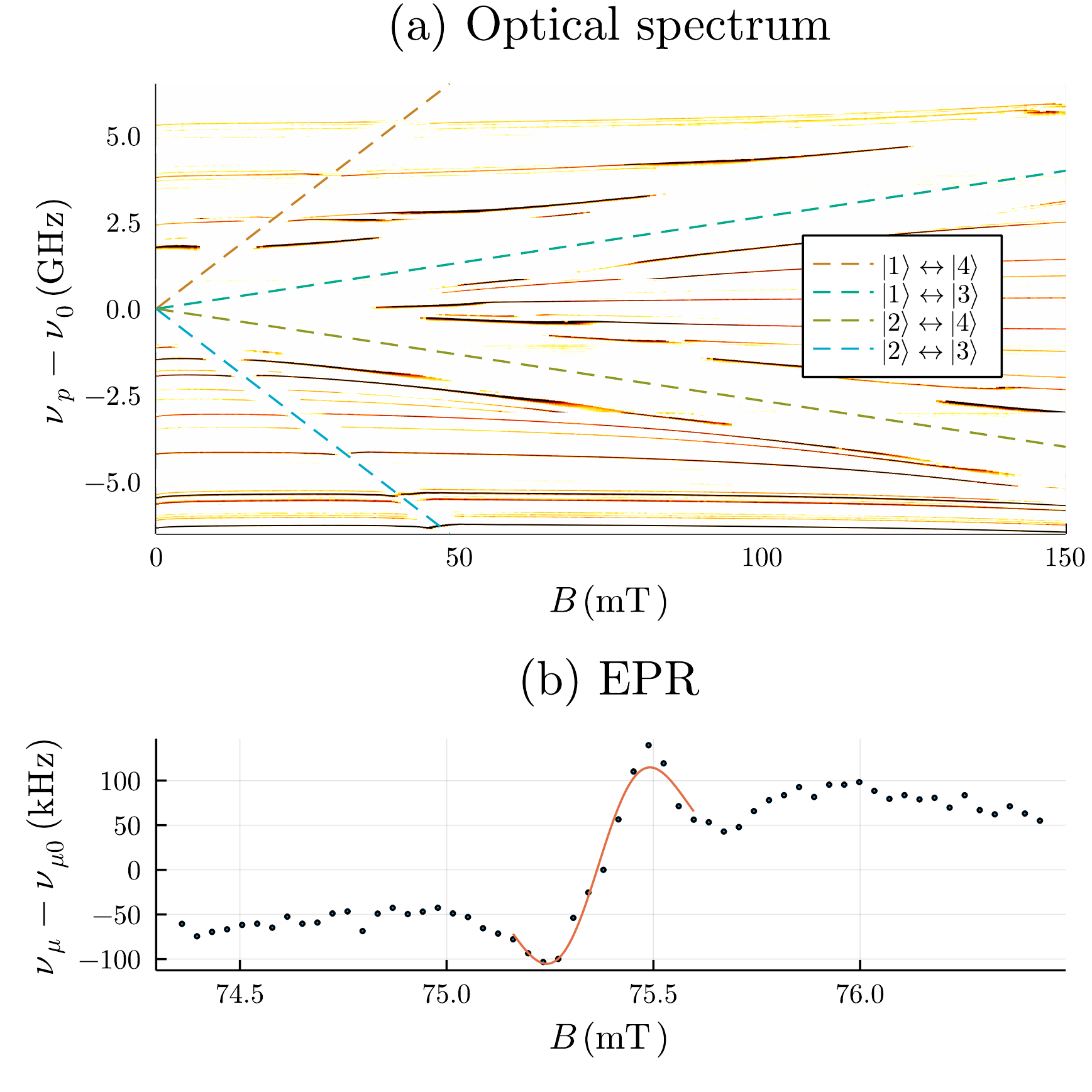}
    \caption{
    (a) TE-polarized optical modes in the Er\textsuperscript{3+}:YSO resonator as a function of magnetic field. $\nu_0$ is the zero-field ${}^4I_{13/2}\leftrightarrow{}^4I_{15/2}$ transition frequency.
    These modes are measured by the reflection from the coupling prism; dark lines show where light has coupled into the cavity.
    To increase visibility of the narrow modes on a large frequency scale, the reflection spectra are convolved with a \SI{30}{\mega\hertz} wide top-hat function.
    Dashed lines show the optical transitions of the erbium ions, from which we can estimate the effective $g$-factors of the ground and excited state are $g_{g,\mathrm{eff}} \approx 11.5$, and $g_{e,\mathrm{eff}} \approx 7.7$. These are about \SI{10}{\percent} higher than the values measured along the $D_2$ axis by \textcite{sun_magnetic_2008}.
    Some stitching errors can be seen, as a range of only $\sim\SI{2.2}{\giga\hertz}$ was measured at a time. Slight increases in frequency of the modes near zero applied magnetic field are artefacts of the laser frequency settling over a long time period.
    (b)
    EPR measurement of the microwave coupling via a shift in the microwave mode frequency.
    The fit is the derivative of a Gaussian distribution, from which we find the full-width at half-maximum of the transition, which is \SI{47.4\pm3.5}{\mega\hertz}.
    From the EPR measurement we find $g_{g,\mathrm{eff}} = \num{11.52\pm0.25}$.
    }
    \label{fig:modesplitting}
\end{figure}

To measure the coupling between the microwave resonator and the \Erion{} spin transition, we use an EPR technique based on the Pound locking techniques \cite{pound_electronic_1946}.
The microwave source couples a frequency-modulated microwave tone into the cavity by the input coaxial pin mounted at the top of the microwave cavity.
A microwave power detector rectifies the microwave output from another pin, and a lock-in amplifier detects the modulation thereon which gives an error signal.
Measurement of the error signal at different microwave carrier frequencies provides us with a reference to convert between error signal level and microwave frequency detuning.

Next the microwave carrier frequency is fixed close to the cavity frequency.
When the Zeeman splitting of the ground state approaches resonance with the microwave cavity, the cavity resonant frequency is pulled by the ions causing the error signal to shift, which we can convert into a frequency excursion, as shown in Fig.~\ref{fig:modesplitting}(b). As we expect the \Erion{} spectral density to be a Gaussian distribution, we fit the response to the derivative of a Gaussian, from which we obtain measures of the inhomogeneous linewidth and the total peak-to-peak shift of the microwave resonant frequency.

Prior to the EPR measurement, we first oriented the magnetic field with respect to the YSO crystal axes, as the resonator and magnetic field coils will not be perfectly aligned. Site 1 of \Erion{} in YSO has two distinct orientations related by $C_2$ symmetry \cite{sun_magnetic_2008}. By applying orthogonal magnetic fields $B_y$ and $B_x$ we could tune it such that the two orientations have equal effective $g$ factors, as determined by EPR.
This finds a nearby direction in the $D_1$--$D_2$ plane.

The novel measurement technique we introduce here is a variation on ODMR, possible due to the strong optical coupling between the WGMs and the collective erbium transitions.
The PDH error signal is integrated by an open-software closed-hardware FPGA board \cite{neuhaus_pyrpl_2017}, which feeds back into the laser's internal piezo to lock to the WGM.
The microwave resonator is excited; if the driving frequency matches the Zeeman splitting of \Erion{}, the optical mode will shift and be detected on the PDH error signal by the lock-in amplifier.
The input microwave field is amplitude modulated with a modulation depth of \SI{45}{\percent} and a modulation frequency of \SI{1}{\kilo\hertz}. The lock-in amplifier detects amplitude modulation of the error signal at this frequency.
Dual use of the error signal for both locking and detection is similar to the AC-magnetostrictive sensing performed in Ref.~\cite{yu_optomechanical_2016}.
The lock-in amplifier signal is converted into a frequency shift based on the measured slope of the error signal around the optical resonant frequency.
The servo applies a \SI{151.8}{\hertz} digital low-pass filter to its input to avoid the lock compensating for the modulation and thus reducing the signal measured by the lock-in amplifier.
In contrast to regular ODMR where the population transfer is measured by changes to an absorption coefficient, we measure changes in coupling between our WGMs and an optical-frequency \Erion{} transition.

\section{Results}

\begin{figure*}
    \subfloat[$P_p=\SI{2.3}{\milli\watt}$, $T\approx\SI{4}{\kelvin}$]{\includegraphics[width=0.3077\textwidth]{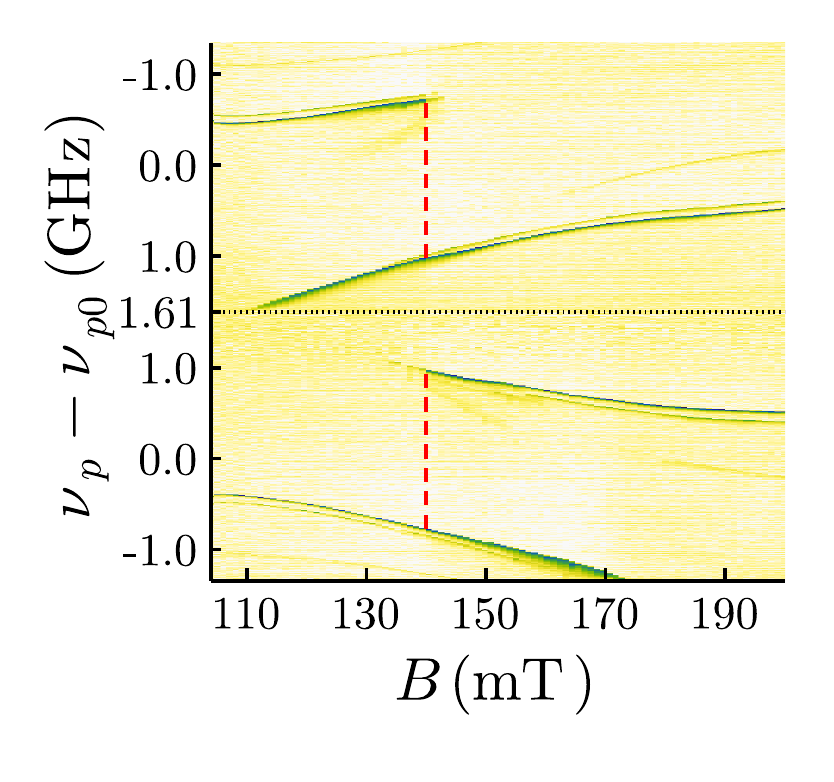}}
    \subfloat[$P_p=\SI{15}{\milli\watt}$, $T\approx\SI{4}{\kelvin}$]{\includegraphics[width=0.3077\textwidth]{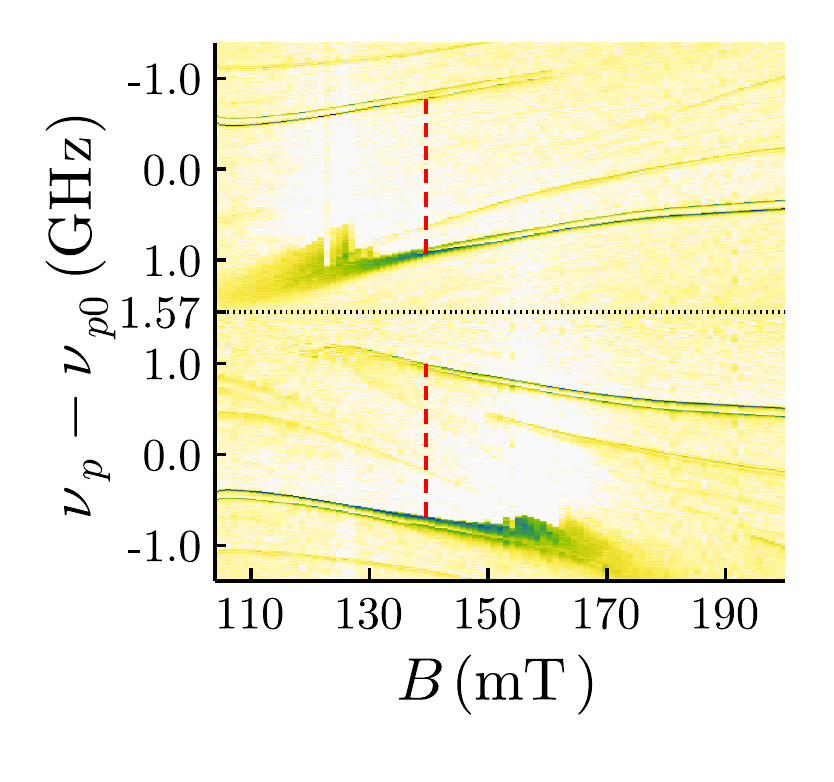}}
    \subfloat[$P_p=\SI{15}{\milli\watt}$, $T\approx\SI{2.9}{\kelvin}${\kern0.0769\textwidth}]{\includegraphics[width=0.3846\textwidth]{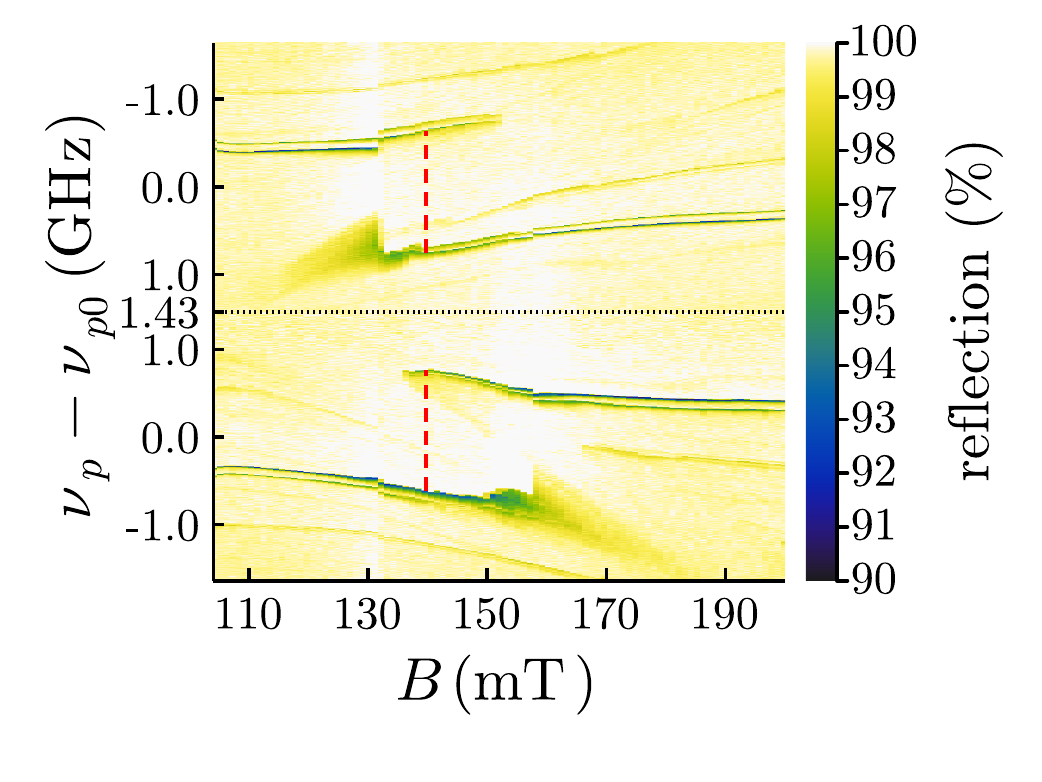}}
    \caption{
        Saturation of optical anticrossing with input optical power. This TM mode has a zero-field frequency of $\nu_{p0}=\SI{195112.8\pm0.4}{\giga\hertz}$. The reflection of the optical resonator is measured over a range of about \SI{3}{\giga\hertz} at different magnetic fields along the $D_2$ axis. The frequency is swept by a triangle wave input to the lasers piezo from the bottom to the top of the graph. Both sides of the frequency sweep are shown to display the asymmetry.
    }
    \label{fig:opticalsat}
\end{figure*}

Optically, we can measure the coupling to the erbium ions by observing the crossings. Fig.~\ref{fig:opticalsat}(a) shows the avoided crossing of two neighboring TM modes separated by about \SI{80}{\mega\hertz}. The better coupled mode has a frequency of $\nu_{p0}=\SI{195112.8\pm0.4}{\giga\hertz}$ at zero applied magnetic field. This mode is chosen as it is one of the few TM modes that couples well. The optical power is $P_p=\SI{2.3}{\milli\watt}$, and the maximum coupling efficiency to this mode is \SI{10}{\percent}. 
The input laser frequency is scanned in a sawtooth. 
Sweeping the frequency up gives a different line shape to sweeping in the opposite direction.

To gauge the dispersion induced by the erbium ions, we measure the FSRs of this TM mode family at zero applied magnetic field. The measurements are performed by sideband spectroscopy \cite{li_sideband_2012}. The input laser is swept across the resonance and modulated by the EOM. When the modulated frequency is equal to the FSR, the absorption dips from the carrier coupling to the probed resonance and a sideband coupling to the mode one FSR away coincide. By varying the modulation frequency and examining the transmitted line shapes the FSR can be measured to a resolution better than the linewidth. Seven consecutive FSR measurements are given in Table \ref{tab:FSRs}.
The group velocity dispersion (GVD) $\beta_2$ can be approximated by
\begin{align}
    \beta_2=\frac{\partial^2\beta}{\partial\omega^2}
    &= -\frac{1}{(2\pi)^2}\frac{\partial^2\nu}{\partial\beta^2}\left(\frac{\partial\beta}{\partial\nu}\right)^3 
    \nonumber
    \\
    &\approx -\frac{1}{R(2\pi)^2}\frac{\nu_{m+1}-2\nu_m+\nu_{m-1}}{(\nu_{m+1}/2+\nu_{m-1}/2)^3},
\end{align}
where $\beta=m/R$ is the propagation constant.
The value of GVD with the largest magnitude is estimated to be $\beta_2 \approx \SI{-4.6e9}{\square\femto\second\per\meter}$.

\begin{table}
    \caption{Free spectral ranges (FSRs) of the TM mode family with a mode which has an azimuthal mode number of $m_\mathrm{ref}\approx\num{15000}$ and a frequency of $\nu_{m_\mathrm{ref}} = \SI{195112.8\pm0.4}{\giga\hertz}$. This is the better coupled of the pair of modes shown in in Fig.~\ref{fig:opticalsat}.
    The right column gives the difference between adjacent FSRs, which gives a measure of dispersion.
    All FSRs carry a measurement uncertainty of \SI{0.5}{\mega\hertz}.
    }
    \begin{ruledtabular}
    \begin{tabular}{dcd}
        \multicolumn{1}{c}{Mode number} &
            \multicolumn{1}{c}{FSR (\si{\giga\hertz})} &
            \multicolumn{1}{c}{$\Delta$FSR (\si{\mega\hertz})} \\
        \multicolumn{1}{c}{$m - m_\mathrm{ref}$} & 
            \multicolumn{1}{c}{$\nu_{m+1} - \nu_m$}  & 
            \multicolumn{1}{c}{$\nu_{m+1}-2\nu_m+\nu_{m-1}$} \\ \hline
        3 & 12.3255 & 13.0 \\ 
        2 & 12.3125 & 65.5 \\ 
        1 & 12.2470 & -517.0  \\ 
        0 & 12.8180 & 731.0 \\ 
        -1 & 12.0870 & -213.0  \\ 
        -2 & 12.3000 & -21.5 \\ 
        -3 & 12.3215 & \\
    \end{tabular}
    \end{ruledtabular}
    \label{tab:FSRs}
\end{table}

The optical powers reported are the geometric mean of the powers measured entering and exiting the cryostat when there is no coupling to the optical resonator, $P_p = \sqrt{P_\mathrm{enter}P_\mathrm{exit}}$. If all losses were due to Fresnel reflections, this would accurately give the power at the prism--resonator interface due to the symmetry of the optical path in the cryostat. However---as a small amount of clipping of the beam is observed---this power will be a slight overestimate.

By increasing the optical power, the splitting of the avoided crossing decreases due to saturation. Increasing the power to \SI{15}{\milli\watt} decreases the splitting from \SI{1.8}{\giga\hertz} to \SI{1.7}{\giga\hertz}.
A further decrease is seen by lowering the temperature of the cryostat. Fig.~\ref{fig:opticalsat} shows the same mode measured with \SI{15}{\milli\watt} of input power with a lowered base temperature of about \SIrange{2.8}{2.9}{\kelvin} as measured on the exterior of the microwave cavity before and after measurement. The jumps at about \SIlist{132;158}{\milli\tesla} show the significant decrease in optical pumping occurring between these two values.

\begin{figure}[b]
    \includegraphics[width=1\columnwidth]{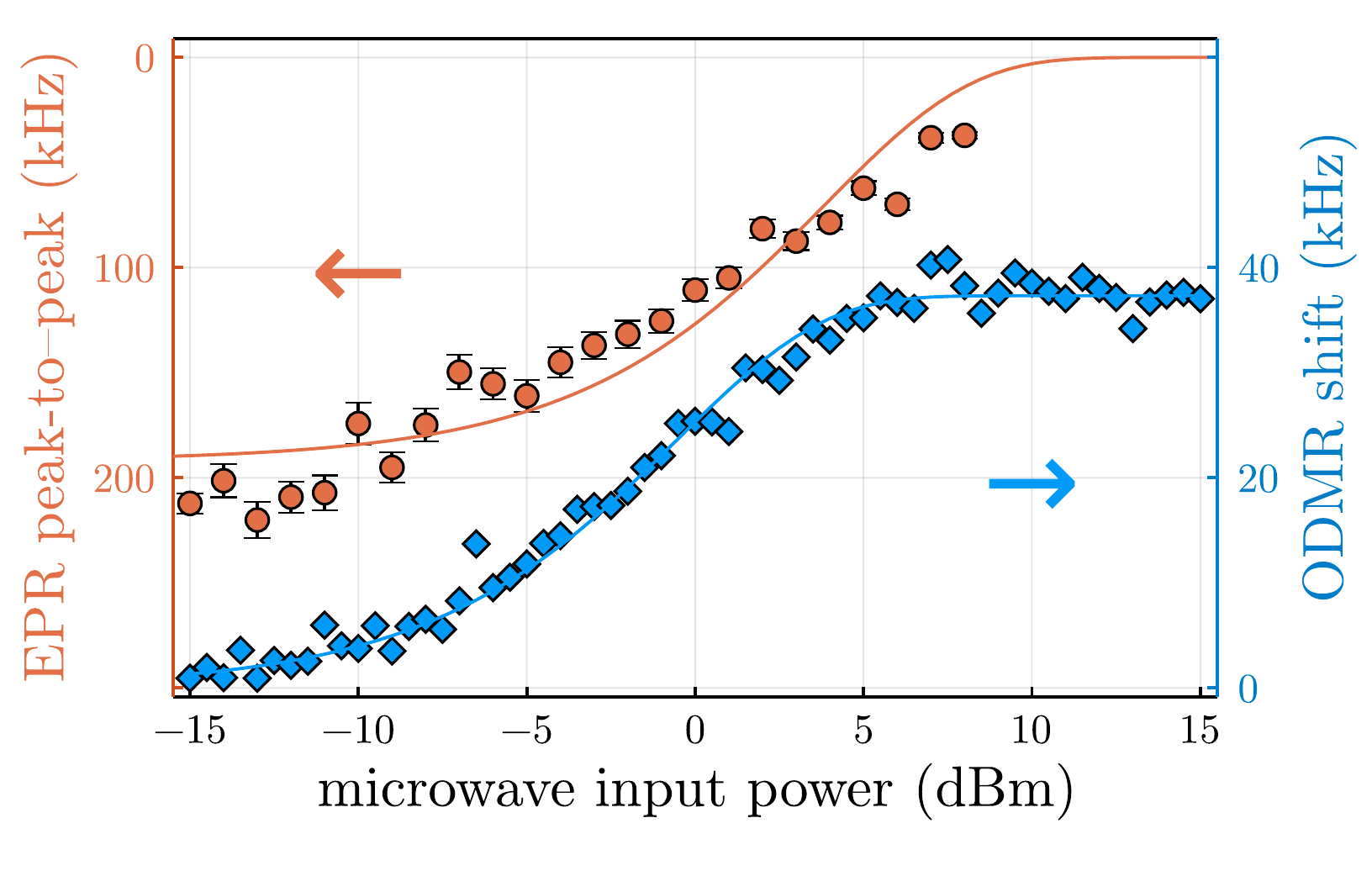}
    \caption{
        Saturation of EPR and ODMR.
        EPR microwave mode shift (orange circles, left axis) saturates towards zero as the ground-state population imbalance decreases with pumping.
        ODMR optical mode shift (blue diamonds, right axis) saturates to the microwave mode pump moving a maximum population difference.
        The optical mode probed is a TE mode at \SI{195106.1\pm0.4}{\giga\hertz}, slightly lower in frequency than the one used for Fig.~\ref{fig:ODMR}(a).
        The fitted $(1/e)$ saturation powers are \SI{3.82\pm0.50}{dBm} for EPR and \SI{-0.54\pm0.15}{dBm} for ODMR.
    }
    \label{fig:ODMRsat}
\end{figure}

We characterize the microwave transition with EPR. Fig.~\ref{fig:modesplitting}(b) shows a measured EPR response with a fit showing the peak-to-peak frequency shift of the microwave mode and Gaussian full width at half maximum (FWHM). Fig.~\ref{fig:ODMRsat} shows the frequency shift at different powers, which decreases as the spin transition saturates.
The microwave powers reported are the powers entering the cryostat; the power coupled into the microwave resonator is about 40\,dB lower.
Throughout these measurements the fitted FWHM stays relatively constant. We find a FWHM of \SI{47.8}{\mega\hertz}, with an error of $\pm\SI{0.7}{\mega\hertz}$ derived from the statistical uncertainty of the fits assuming they are independent, and an error of $\pm\SI{0.6}{\mega\hertz}$ from our systematic uncertainty in the magnetic field hysteresis, which is common to all measurements carried out during the same magnet cool-down.

\begin{figure*}
    \subfloat[][]{\includegraphics[width=0.33\textwidth]{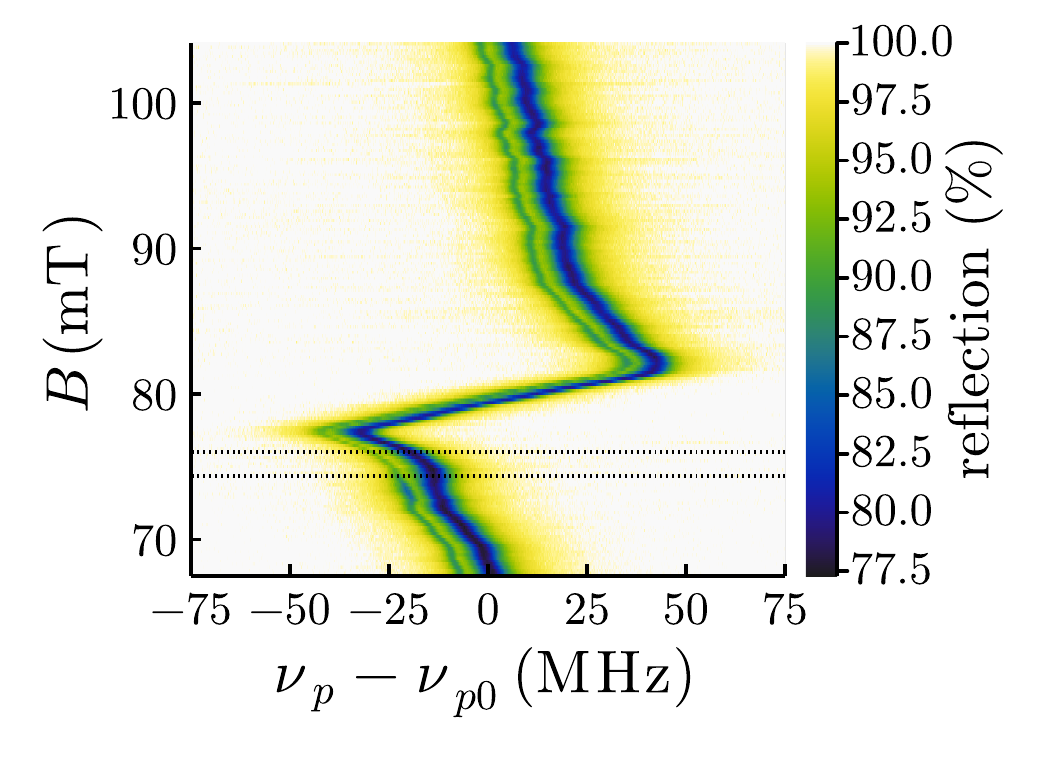}}
    \subfloat[][]{\includegraphics[width=0.33\textwidth]{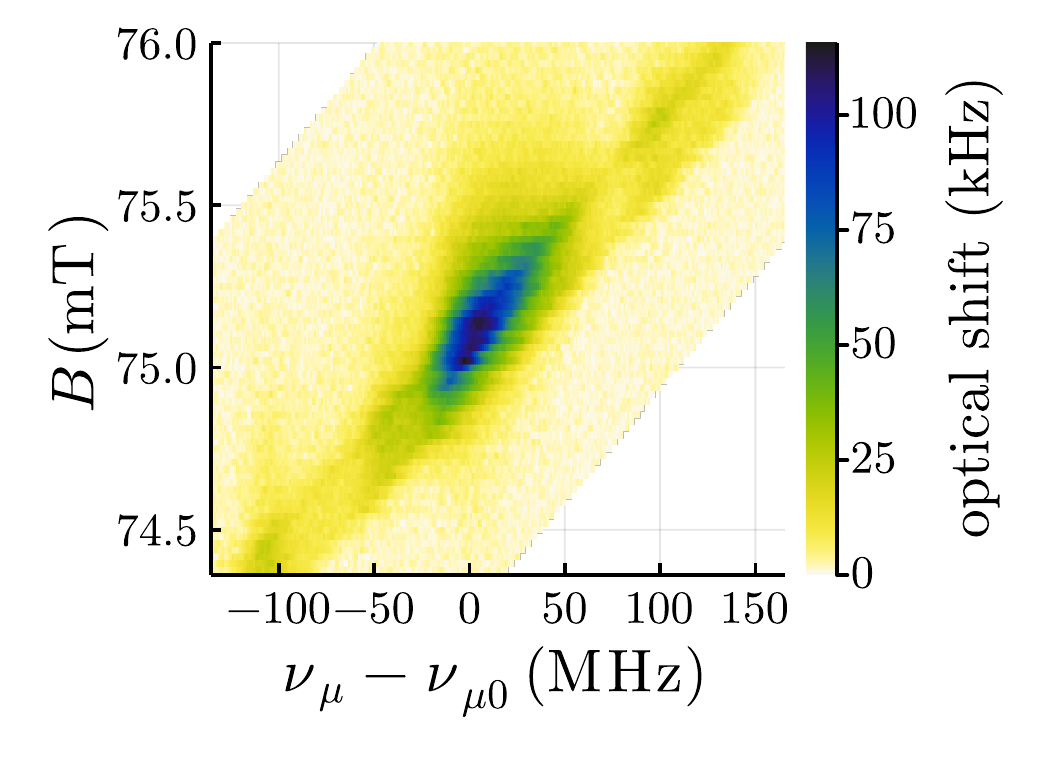}}
    \subfloat[][]{\includegraphics[width=0.33\textwidth]{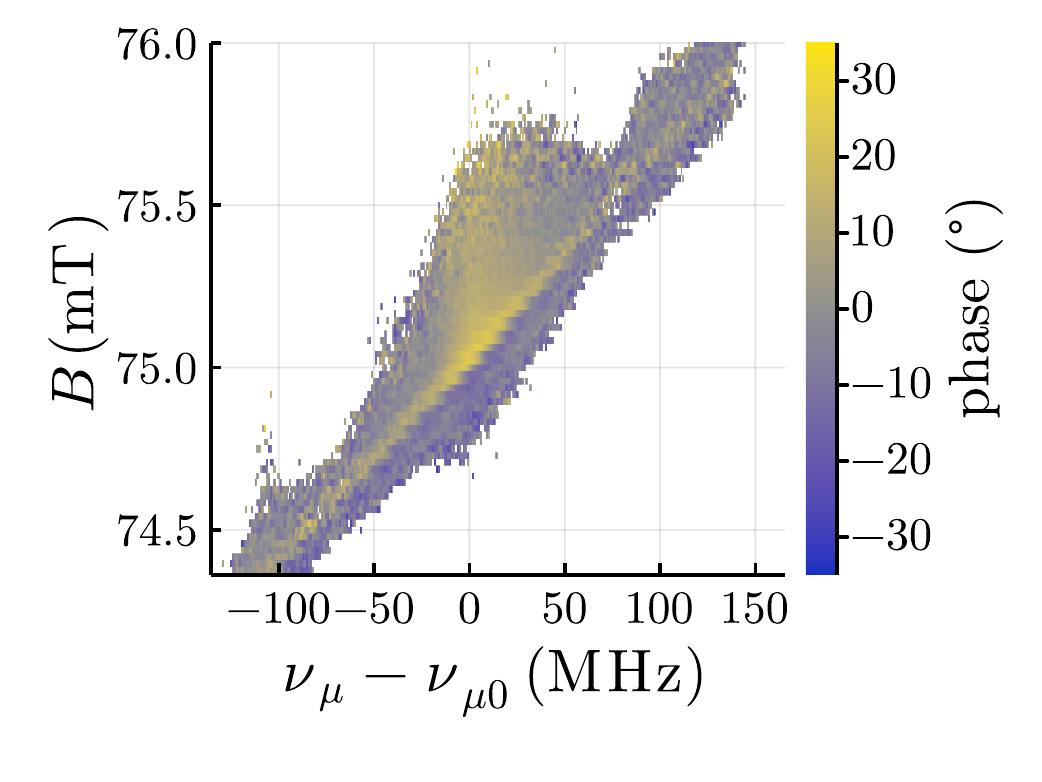}}
    \\ \vspace{-1em}
    \subfloat[][]{\includegraphics[width=0.33\textwidth]{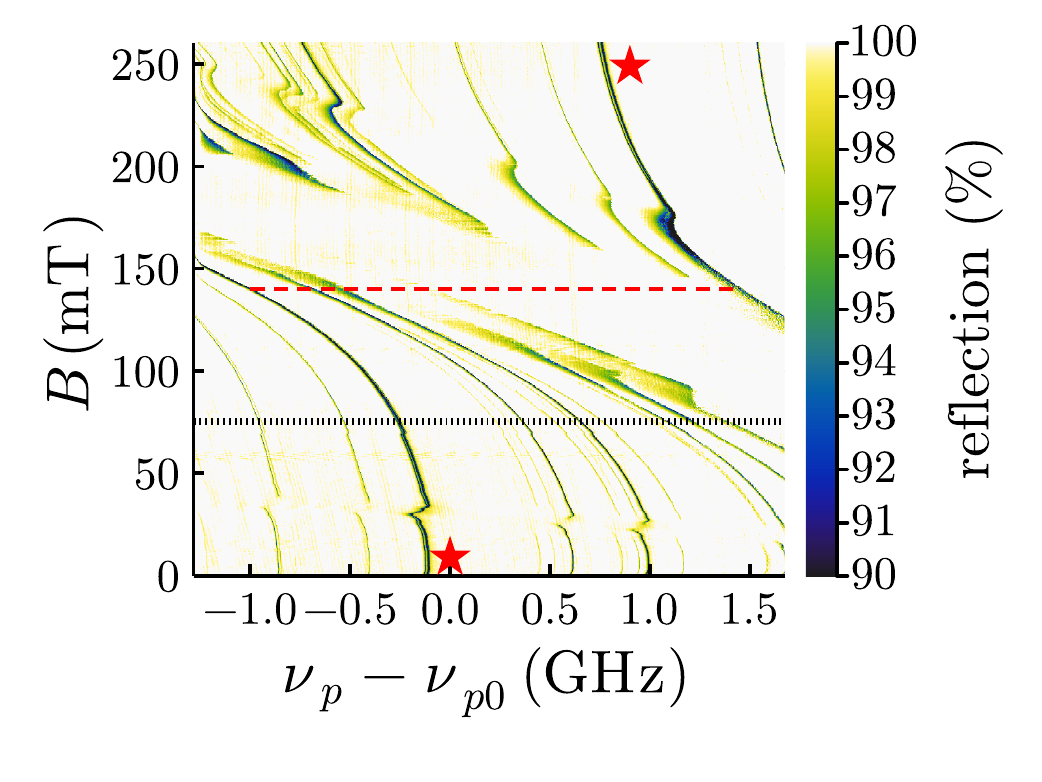}}
    \subfloat[][]{\includegraphics[width=0.33\textwidth]{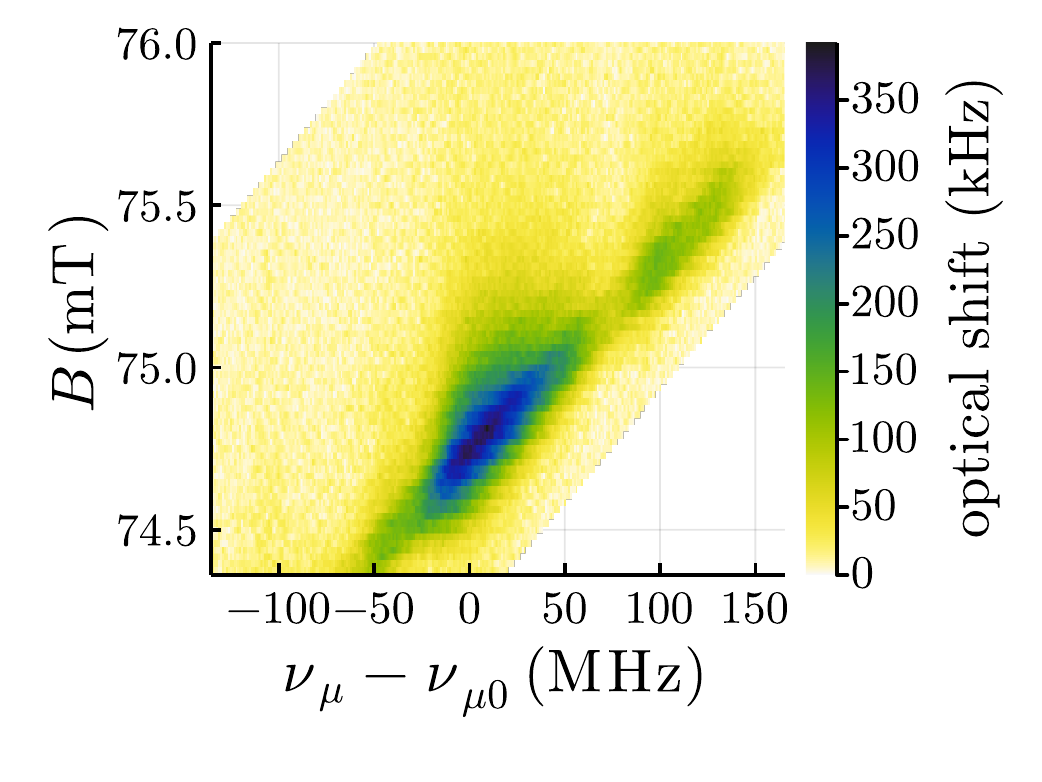}}
    \subfloat[][]{\includegraphics[width=0.33\textwidth]{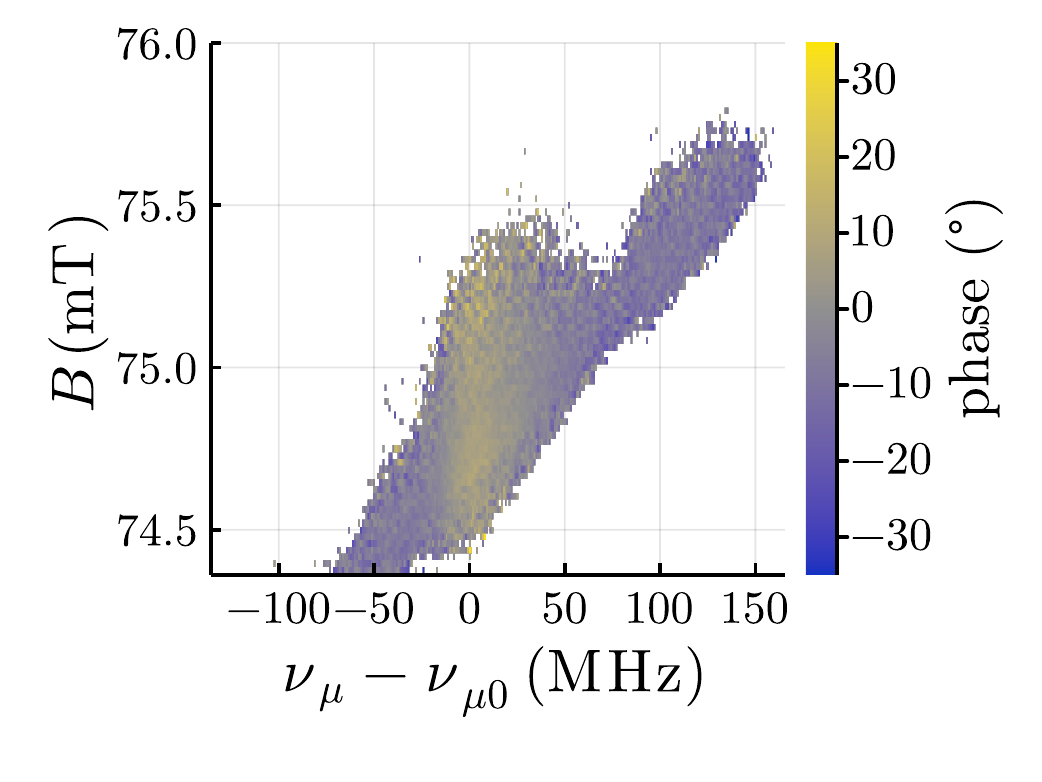}}
    \caption{
        (a) Frequency shift of a TE optical mode with magnetic field near a frequency of $\nu_{p0} = \SI{195106.2\pm0.4}{\giga\hertz}$.
        At a magnetic field of about \SI{81}{\milli\tesla} it crosses the $\lvert 2\rangle \leftrightarrow \lvert 3\rangle$ transition.
        (d) Frequency shift of a TE optical mode with magnetic field near a frequency of $\nu_{p0} = \SI{195112.7\pm0.4}{\giga\hertz}$.
        Between about \SIlist{100;200}{\milli\tesla} it avoids the $\lvert 2\rangle \leftrightarrow \lvert 4\rangle$ transition. Red stars at the top and bottom are located just beside the mode of interest, and a red dashed line shows the mode splitting of \SI{2.45}{\giga\hertz}.
        Black dotted lines in (a,d) indicate the magnetic field strengths examined in the other parts of this figure.
        (b,e) Frequency shifts of the optical mode shown in (a,d) when a magnetic field is applied along the $D_2$ axis and the microwave cavity is pumped at different frequencies. The diagonal line shows the Zeeman splitting of the \Erion{}. Where this splitting equals the microwave resonance frequency of $\nu_{\mu0}=\SI{12.155}{\giga\hertz}$ the maximum shift is attained.
        The magnetic field values are converted directly from the currents set on the current sources. Hysteresis in the superconducting magnets causes an offset in current, and hence an offset in assumed magnetic fields between the two measurements: (a-c) and (d-f).
        (c,f) Relative phases of the response with respect to the modulating \SI{1}{\kilo\hertz}.
        The microwave input power into the cryostat for (b,c,e,f) is 5\,dBm, which is amplitude modulated at \SI{1}{\kilo\hertz} with a modulation depth of \SI{45}{\percent}.
    }
    \label{fig:ODMR}
\end{figure*}

Next we examine the ODMR frequency shifts. We look at two sets of TE optical modes in different polariton regimes. Fig.~\ref{fig:ODMR}(a) shows the first pair of modes with a frequency of $\nu_{p0} = \SI{195106.2\pm0.4}{\giga\hertz}$. At a magnetic field of about \SI{81}{\milli\tesla} they cross the $\lvert 2\rangle \leftrightarrow \lvert 3\rangle$ transition.  Fig.~\ref{fig:ODMR}(d) shows the second group of three modes with a frequency of $\nu_{p0} = \SI{195112.7\pm0.4}{\giga\hertz}$. Between about \SIlist{100;200}{\milli\tesla} they avoid the $\lvert 1\rangle \leftrightarrow \lvert 3\rangle$ transition. The splitting of \SI{2.4}{\giga\hertz} reveals a coupling strength of $2\pi\times\SI{1.2}{\giga\hertz}$. Red stars on the top and bottom are located just beside the modes of interest. For clarity the colormap is throttled at \SI{90}{\percent}; the maximum coupling obtained for this mode is about \SI{25}{\percent}. To simplify the error signal and to allow it to measure larger frequency shifts, we overcouple to these modes in later experiments such that the error signal is monotonic around the resonance frequency.

Fig.~\ref{fig:ODMR}(b,e) show the measured frequency shifts of these modes. At each magnetic field value the microwave pump frequency $\nu_\mu$ is slowly increased. Where the ground-state splitting of the erbium ions equals the microwave resonance frequency of \SI{12.155}{\giga\hertz} the maximum shift is attained. The response is asymmetric, especially in Fig.~\ref{fig:ODMR}(b), which is probably due to optical hole burning decreasing the response for frequencies above the Zeeman splitting.
The phase response of these shifts is shown in Figs.~\ref{fig:ODMR}(c,f) for measurements above the noise floor. The phase response of Fig.~\ref{fig:ODMR}(c) further accentuates the asymmetry we attribute to optical hole burning for this optical frequency, which is closer to a bare erbium transition at these magnetic field values.

The ODMR frequency shift of the optical mode saturates at high microwave power, as shown in Fig.~\ref{fig:ODMRsat}.
\begin{equation}
\label{eq:ODMRsat}
    \Delta \nu(P) = \Delta\nu_\mathrm{max} \left[ 1 - \exp\left(-P/P_\mathrm{sat}\right) \right],
\end{equation}
where $\Delta\nu_\mathrm{max}$ is the maximum attainable frequency shift and $P_\mathrm{sat}$ is the saturation microwave power.

\section{Discussion}

We have shown that high $Q$-factors can be obtained in cryogenic rare-earth-ion-doped whispering-gallery resonators, even when the modes are very close to optical transitions.
The WGMs observed couple strongly to ensemble excitations of the \SI{50}{ppm} erbium doping.

The modified ODMR spectroscopy we have described here is an elaboration of the technique to probe the spins with an optical resonance that couples to the optical transition stronger than its linewidth. We have shown that we can measure the ODMR response with a polariton mode in the regime where it is mostly photon like.
The saturation microwave power for ODMR is smaller than that of EPR (\SI{-0.54\pm0.15}{dBm} vs.~\SI{3.82\pm0.50}{dBm}). This shows that the microwave magnetic field is high in the optical mode volume, as the ODMR only probes the subclass of ions that live therein.

In the future we aim to observe Raman heterodyne \cite{mlynek_raman_1983} in the resonator. We have tried to observe Raman heterodyne in the system but could not. 
We believe this is due to the large imbalance in branching ratio for different erbium transitions\footnote{{cf}. coupling in Figs.~\ref{fig:ODMR}(a) and \ref{fig:ODMR}(d)} making the effective FSR significantly different from the microwave cavity frequency.
The branching ratio between spin states can be optimized by careful choice of static magnetic field direction \cite{fernandez-gonzalvo_coherent_2015,king_probing_2021}; such an orientation is not accessible in our current experimental configuration.
Such Raman heterodyne is a stepping stone for coherent microwave--optical transduction, where adiabatic elimination of the excited states allows the process to have a high fidelity \cite{fernandez-gonzalvo_cavity-enhanced_2019,williamson_magneto-optic_2014}.

In conclusion, we report an erbium-doped WGM resonator sitting inside a microwave cavity for a modified ODMR measurement.
The high $Q$-factor optical modes show strong coupling to ensemble erbium transitions.
Our modified ODMR spectroscopy measures spin state through the changes in this coupling strength.
We can therefore detect the spin transitions with an optical probe detuned by more than the inhomogeneous linewidth.
Compared to EPR, strong coupling ODMR has the capability of responding to only the ions within the optical mode volume, which can be located in a region of a stronger average microwave field.
Additionally, only those ions could be used for Raman heterodyne and therefore this characterization is important for future investigations to enable microwave--optical transduction.

% \bibliography{WGMODMR}

\begin{thebibliography}{51}%
\makeatletter
\providecommand \@ifxundefined [1]{%
 \@ifx{#1\undefined}
}%
\providecommand \@ifnum [1]{%
 \ifnum #1\expandafter \@firstoftwo
 \else \expandafter \@secondoftwo
 \fi
}%
\providecommand \@ifx [1]{%
 \ifx #1\expandafter \@firstoftwo
 \else \expandafter \@secondoftwo
 \fi
}%
\providecommand \natexlab [1]{#1}%
\providecommand \enquote  [1]{``#1''}%
\providecommand \bibnamefont  [1]{#1}%
\providecommand \bibfnamefont [1]{#1}%
\providecommand \citenamefont [1]{#1}%
\providecommand \href@noop [0]{\@secondoftwo}%
\providecommand \href [0]{\begingroup \@sanitize@url \@href}%
\providecommand \@href[1]{\@@startlink{#1}\@@href}%
\providecommand \@@href[1]{\endgroup#1\@@endlink}%
\providecommand \@sanitize@url [0]{\catcode `\\12\catcode `\$12\catcode
  `\&12\catcode `\#12\catcode `\^12\catcode `\_12\catcode `\%12\relax}%
\providecommand \@@startlink[1]{}%
\providecommand \@@endlink[0]{}%
\providecommand \url  [0]{\begingroup\@sanitize@url \@url }%
\providecommand \@url [1]{\endgroup\@href {#1}{\urlprefix }}%
\providecommand \urlprefix  [0]{URL }%
\providecommand \Eprint [0]{\href }%
\providecommand \doibase [0]{https://doi.org/}%
\providecommand \selectlanguage [0]{\@gobble}%
\providecommand \bibinfo  [0]{\@secondoftwo}%
\providecommand \bibfield  [0]{\@secondoftwo}%
\providecommand \translation [1]{[#1]}%
\providecommand \BibitemOpen [0]{}%
\providecommand \bibitemStop [0]{}%
\providecommand \bibitemNoStop [0]{.\EOS\space}%
\providecommand \EOS [0]{\spacefactor3000\relax}%
\providecommand \BibitemShut  [1]{\csname bibitem#1\endcsname}%
\let\auto@bib@innerbib\@empty
%</preamble>
\bibitem [{\citenamefont {Böttger}\ \emph {et~al.}(2009)\citenamefont
  {Böttger}, \citenamefont {Thiel}, \citenamefont {Cone},\ and\ \citenamefont
  {Sun}}]{bottger_effects_2009}%
  \BibitemOpen
  \bibfield  {author} {\bibinfo {author} {\bibfnamefont {T.}~\bibnamefont
  {B{\"o}ttger}}, \bibinfo {author} {\bibfnamefont {C.~W.}\ \bibnamefont {Thiel}},
  \bibinfo {author} {\bibfnamefont {R.~L.}\ \bibnamefont {Cone}},\ and\
  \bibinfo {author} {\bibfnamefont {Y.}~\bibnamefont {Sun}},\ }\bibfield
  {title} {\bibinfo {title} {Effects of magnetic field orientation on optical
  decoherence in {Er$^{3+}$:Y$_2$SiO$_5$}},\ }\href
  {https://doi.org/10.1103/PhysRevB.79.115104} {\bibfield  {journal} {\bibinfo
  {journal} {Physical Review B}\ }\textbf {\bibinfo {volume} {79}},\ \bibinfo
  {pages} {115104} (\bibinfo {year} {2009})}\BibitemShut {NoStop}%
\bibitem [{\citenamefont {Sun}\ \emph {et~al.}(2002)\citenamefont {Sun},
  \citenamefont {Thiel}, \citenamefont {Cone}, \citenamefont {Equall},\ and\
  \citenamefont {Hutcheson}}]{sun_recent_2002}%
  \BibitemOpen
  \bibfield  {author} {\bibinfo {author} {\bibfnamefont {Y.}~\bibnamefont
  {Sun}}, \bibinfo {author} {\bibfnamefont {C.~W.}\ \bibnamefont {Thiel}},
  \bibinfo {author} {\bibfnamefont {R.~L.}\ \bibnamefont {Cone}}, \bibinfo
  {author} {\bibfnamefont {R.~W.}\ \bibnamefont {Equall}},\ and\ \bibinfo
  {author} {\bibfnamefont {R.~L.}\ \bibnamefont {Hutcheson}},\ }\bibfield
  {title} {\bibinfo {title} {Recent progress in developing new rare earth
  materials for hole burning and coherent transient applications},\ }\href
  {https://doi.org/10.1016/S0022-2313(02)00281-8} {\bibfield  {journal}
  {\bibinfo  {journal} {Journal of Luminescence}\ }\bibinfo {series}
  {Proceedings of the Seventh International Meeting on Hole Burning, Single
  Molecules and Related Spectroscopies: Science and Applications},\ \textbf
  {\bibinfo {volume} {98}},\ \bibinfo {pages} {281} (\bibinfo {year}
  {2002})}\BibitemShut {NoStop}%
\bibitem [{\citenamefont {Equall}\ \emph {et~al.}(1994)\citenamefont {Equall},
  \citenamefont {Sun}, \citenamefont {Cone},\ and\ \citenamefont
  {Macfarlane}}]{equall_ultraslow_1994}%
  \BibitemOpen
  \bibfield  {author} {\bibinfo {author} {\bibfnamefont {R.~W.}\ \bibnamefont
  {Equall}}, \bibinfo {author} {\bibfnamefont {Y.}~\bibnamefont {Sun}},
  \bibinfo {author} {\bibfnamefont {R.~L.}\ \bibnamefont {Cone}},\ and\
  \bibinfo {author} {\bibfnamefont {R.~M.}\ \bibnamefont {Macfarlane}},\
  }\bibfield  {title} {\bibinfo {title} {Ultraslow optical dephasing in
  {Eu$^{3+}$:Y$_2$SiO$_5$}},\ }\href
  {https://doi.org/10.1103/PhysRevLett.72.2179} {\bibfield  {journal} {\bibinfo
   {journal} {Physical Review Letters}\ }\textbf {\bibinfo {volume} {72}},\
  \bibinfo {pages} {2179--2182} (\bibinfo {year} {1994})}\BibitemShut
  {NoStop}%
\bibitem [{\citenamefont {Ahlefeldt}\ \emph {et~al.}(2016)\citenamefont
  {Ahlefeldt}, \citenamefont {Hush},\ and\ \citenamefont
  {Sellars}}]{ahlefeldt_ultranarrow_2016}%
  \BibitemOpen
  \bibfield  {author} {\bibinfo {author} {\bibfnamefont {R.~L.}\ \bibnamefont
  {Ahlefeldt}}, \bibinfo {author} {\bibfnamefont {M.~R.}\ \bibnamefont
  {Hush}},\ and\ \bibinfo {author} {\bibfnamefont {M.~J.}\ \bibnamefont
  {Sellars}},\ }\bibfield  {title} {\bibinfo {title} {Ultranarrow {{Optical
  Inhomogeneous Linewidth}} in a {{Stoichiometric Rare-Earth Crystal}}},\
  }\href {https://doi.org/10.1103/PhysRevLett.117.250504} {\bibfield  {journal}
  {\bibinfo  {journal} {Physical Review Letters}\ }\textbf {\bibinfo {volume}
  {117}},\ \bibinfo {pages} {250504} (\bibinfo {year} {2016})}\BibitemShut
  {NoStop}%
\bibitem [{\citenamefont {Macfarlane}\ \emph {et~al.}(1992)\citenamefont
  {Macfarlane}, \citenamefont {Cassanho},\ and\ \citenamefont
  {Meltzer}}]{macfarlane_inhomogeneous_1992}%
  \BibitemOpen
  \bibfield  {author} {\bibinfo {author} {\bibfnamefont {R.~M.}\ \bibnamefont
  {Macfarlane}}, \bibinfo {author} {\bibfnamefont {A.}~\bibnamefont
  {Cassanho}},\ and\ \bibinfo {author} {\bibfnamefont {R.~S.}\ \bibnamefont
  {Meltzer}},\ }\bibfield  {title} {\bibinfo {title} {Inhomogeneous broadening
  by nuclear spin fields: {{A}} new limit for optical transitions in solids},\
  }\href {https://doi.org/10.1103/PhysRevLett.69.542} {\bibfield  {journal}
  {\bibinfo  {journal} {Physical Review Letters}\ }\textbf {\bibinfo {volume}
  {69}},\ \bibinfo {pages} {542} (\bibinfo {year} {1992})}\BibitemShut
  {NoStop}%
\bibitem [{\citenamefont {Zhong}\ \emph {et~al.}(2015)\citenamefont {Zhong},
  \citenamefont {Hedges}, \citenamefont {Ahlefeldt}, \citenamefont
  {Bartholomew}, \citenamefont {Beavan}, \citenamefont {Wittig}, \citenamefont
  {Longdell},\ and\ \citenamefont {Sellars}}]{zhong_optically_2015}%
  \BibitemOpen
  \bibfield  {author} {\bibinfo {author} {\bibfnamefont {M.}~\bibnamefont
  {Zhong}}, \bibinfo {author} {\bibfnamefont {M.~P.}\ \bibnamefont {Hedges}},
  \bibinfo {author} {\bibfnamefont {R.~L.}\ \bibnamefont {Ahlefeldt}}, \bibinfo
  {author} {\bibfnamefont {J.~G.}\ \bibnamefont {Bartholomew}}, \bibinfo
  {author} {\bibfnamefont {S.~E.}\ \bibnamefont {Beavan}}, \bibinfo {author}
  {\bibfnamefont {S.~M.}\ \bibnamefont {Wittig}}, \bibinfo {author}
  {\bibfnamefont {J.~J.}\ \bibnamefont {Longdell}},\ and\ \bibinfo {author}
  {\bibfnamefont {M.~J.}\ \bibnamefont {Sellars}},\ }\bibfield  {title}
  {\bibinfo {title} {Optically addressable nuclear spins in a solid with a
  six-hour coherence time},\ }\href {https://doi.org/10.1038/nature14025}
  {\bibfield  {journal} {\bibinfo  {journal} {Nature}\ }\textbf {\bibinfo
  {volume} {517}},\ \bibinfo {pages} {177} (\bibinfo {year}
  {2015})}\BibitemShut {NoStop}%
\bibitem [{\citenamefont {Ortu}\ \emph {et~al.}(2018)\citenamefont {Ortu},
  \citenamefont {Tiranov}, \citenamefont {Welinski}, \citenamefont
  {Fr{\"o}wis}, \citenamefont {Gisin}, \citenamefont {Ferrier}, \citenamefont
  {Goldner},\ and\ \citenamefont {Afzelius}}]{ortu_simultaneous_2018}%
  \BibitemOpen
  \bibfield  {author} {\bibinfo {author} {\bibfnamefont {A.}~\bibnamefont
  {Ortu}}, \bibinfo {author} {\bibfnamefont {A.}~\bibnamefont {Tiranov}},
  \bibinfo {author} {\bibfnamefont {S.}~\bibnamefont {Welinski}}, \bibinfo
  {author} {\bibfnamefont {F.}~\bibnamefont {Fr{\"o}wis}}, \bibinfo {author}
  {\bibfnamefont {N.}~\bibnamefont {Gisin}}, \bibinfo {author} {\bibfnamefont
  {A.}~\bibnamefont {Ferrier}}, \bibinfo {author} {\bibfnamefont
  {P.}~\bibnamefont {Goldner}},\ and\ \bibinfo {author} {\bibfnamefont
  {M.}~\bibnamefont {Afzelius}},\ }\bibfield  {title} {\bibinfo {title}
  {Simultaneous coherence enhancement of optical and microwave transitions in
  solid-state electronic spins},\ }\href
  {https://doi.org/10.1038/s41563-018-0138-x} {\bibfield  {journal} {\bibinfo
  {journal} {Nature Materials}\ }\textbf {\bibinfo {volume} {17}},\ \bibinfo
  {pages} {671} (\bibinfo {year} {2018})}\BibitemShut {NoStop}%
\bibitem [{\citenamefont {Rakonjac}\ \emph {et~al.}(2020)\citenamefont
  {Rakonjac}, \citenamefont {Chen}, \citenamefont {Horvath},\ and\
  \citenamefont {Longdell}}]{rakonjac_long_2020}%
  \BibitemOpen
  \bibfield  {author} {\bibinfo {author} {\bibfnamefont {J.~V.}\ \bibnamefont
  {Rakonjac}}, \bibinfo {author} {\bibfnamefont {Y.-H.}\ \bibnamefont {Chen}},
  \bibinfo {author} {\bibfnamefont {S.~P.}\ \bibnamefont {Horvath}},\ and\
  \bibinfo {author} {\bibfnamefont {J.~J.}\ \bibnamefont {Longdell}},\
  }\bibfield  {title} {\bibinfo {title} {Long spin coherence times in the
  ground state and in an optically excited state of {Er$^{3+}$:Y$_2$SiO$_5$} at
  zero magnetic field},\ }\href {https://doi.org/10.1103/PhysRevB.101.184430}
  {\bibfield  {journal} {\bibinfo  {journal} {Physical Review B}\ }\textbf
  {\bibinfo {volume} {101}},\ \bibinfo {pages} {184430} (\bibinfo {year}
  {2020})}\BibitemShut {NoStop}%
\bibitem [{\citenamefont {Berrington}\ \emph {et~al.}(2022)\citenamefont
  {Berrington}, \citenamefont {R{\o}nnow}, \citenamefont {Sellars},\ and\
  \citenamefont {Ahlefeldt}}]{berrington_negative_2022}%
  \BibitemOpen
  \bibfield  {author} {\bibinfo {author} {\bibfnamefont {M.~C.}\ \bibnamefont
  {Berrington}}, \bibinfo {author} {\bibfnamefont {H.~M.}\ \bibnamefont
  {R{\o}nnow}}, \bibinfo {author} {\bibfnamefont {M.~J.}\ \bibnamefont
  {Sellars}},\ and\ \bibinfo {author} {\bibfnamefont {R.~L.}\ \bibnamefont
  {Ahlefeldt}},\ }\href@noop {} {\bibinfo {title} {Negative refractive index in
  dielectric crystals containing stoichiometric rare-earth ions}} (\bibinfo
  {year} {2022}),\ \Eprint {https://arxiv.org/abs/2205.02739} {arXiv:2205.02739
  [physics.optics]} \BibitemShut {NoStop}%
\bibitem [{\citenamefont {{Lago-Rivera}}\ \emph {et~al.}(2021)\citenamefont
  {{Lago-Rivera}}, \citenamefont {Grandi}, \citenamefont {Rakonjac},
  \citenamefont {Seri},\ and\ \citenamefont {{de
  Riedmatten}}}]{lago-rivera_telecom-heralded_2021}%
  \BibitemOpen
  \bibfield  {author} {\bibinfo {author} {\bibfnamefont {D.}~\bibnamefont
  {{Lago-Rivera}}}, \bibinfo {author} {\bibfnamefont {S.}~\bibnamefont
  {Grandi}}, \bibinfo {author} {\bibfnamefont {J.~V.}\ \bibnamefont
  {Rakonjac}}, \bibinfo {author} {\bibfnamefont {A.}~\bibnamefont {Seri}},\
  and\ \bibinfo {author} {\bibfnamefont {H.}~\bibnamefont {{de Riedmatten}}},\
  }\bibfield  {title} {\bibinfo {title} {Telecom-heralded entanglement between
  multimode solid-state quantum memories},\ }\href
  {https://doi.org/10.1038/s41586-021-03481-8} {\bibfield  {journal} {\bibinfo
  {journal} {Nature}\ }\textbf {\bibinfo {volume} {594}},\ \bibinfo {pages}
  {37} (\bibinfo {year} {2021})}\BibitemShut {NoStop}%
\bibitem [{\citenamefont {Harris}\ \emph {et~al.}(2006)\citenamefont {Harris},
  \citenamefont {Merkel}, \citenamefont {Mohan}, \citenamefont {Chang},
  \citenamefont {Cole}, \citenamefont {Olson},\ and\ \citenamefont
  {Babbitt}}]{harris_multigigahertz_2006}%
  \BibitemOpen
  \bibfield  {author} {\bibinfo {author} {\bibfnamefont {T.~L.}\ \bibnamefont
  {Harris}}, \bibinfo {author} {\bibfnamefont {K.~D.}\ \bibnamefont {Merkel}},
  \bibinfo {author} {\bibfnamefont {R.~K.}\ \bibnamefont {Mohan}}, \bibinfo
  {author} {\bibfnamefont {T.}~\bibnamefont {Chang}}, \bibinfo {author}
  {\bibfnamefont {Z.}~\bibnamefont {Cole}}, \bibinfo {author} {\bibfnamefont
  {A.}~\bibnamefont {Olson}},\ and\ \bibinfo {author} {\bibfnamefont {W.~R.}\
  \bibnamefont {Babbitt}},\ }\bibfield  {title} {\bibinfo {title}
  {Multigigahertz range-{{Doppler}} correlative signal processing in optical
  memory crystals},\ }\href {https://doi.org/10.1364/AO.45.000343} {\bibfield
  {journal} {\bibinfo  {journal} {Applied Optics}\ }\textbf {\bibinfo {volume}
  {45}},\ \bibinfo {pages} {343} (\bibinfo {year} {2006})}\BibitemShut
  {NoStop}%
\bibitem [{\citenamefont {Berger}\ \emph {et~al.}(2016)\citenamefont {Berger},
  \citenamefont {Attal}, \citenamefont {Schwarz}, \citenamefont {Molin},
  \citenamefont {{Louchet-Chauvet}}, \citenamefont {Chaneli{\`e}re},
  \citenamefont {Gou{\"e}t}, \citenamefont {Dolfi},\ and\ \citenamefont
  {Morvan}}]{berger_rf_2016}%
  \BibitemOpen
  \bibfield  {author} {\bibinfo {author} {\bibfnamefont {P.}~\bibnamefont
  {Berger}}, \bibinfo {author} {\bibfnamefont {Y.}~\bibnamefont {Attal}},
  \bibinfo {author} {\bibfnamefont {M.}~\bibnamefont {Schwarz}}, \bibinfo
  {author} {\bibfnamefont {S.}~\bibnamefont {Molin}}, \bibinfo {author}
  {\bibfnamefont {A.}~\bibnamefont {{Louchet-Chauvet}}}, \bibinfo {author}
  {\bibfnamefont {T.}~\bibnamefont {Chaneli{\`e}re}}, \bibinfo {author}
  {\bibfnamefont {J.-L.~L.}\ \bibnamefont {Gou{\"e}t}}, \bibinfo {author}
  {\bibfnamefont {D.}~\bibnamefont {Dolfi}},\ and\ \bibinfo {author}
  {\bibfnamefont {L.}~\bibnamefont {Morvan}},\ }\bibfield  {title} {\bibinfo
  {title} {{{RF Spectrum Analyzer}} for {{Pulsed Signals}}: {{Ultra-Wide
  Instantaneous Bandwidth}}, {{High Sensitivity}}, and {{High
  Time-Resolution}}},\ }\href {https://doi.org/10.1109/JLT.2016.2556008}
  {\bibfield  {journal} {\bibinfo  {journal} {Journal of Lightwave Technology}\
  }\textbf {\bibinfo {volume} {34}},\ \bibinfo {pages} {4658} (\bibinfo {year}
  {2016})}\BibitemShut {NoStop}%
\bibitem [{\citenamefont {Kolesov}\ \emph {et~al.}(2012)\citenamefont
  {Kolesov}, \citenamefont {Xia}, \citenamefont {Reuter}, \citenamefont
  {St{\"o}hr}, \citenamefont {Zappe}, \citenamefont {Meijer}, \citenamefont
  {Hemmer},\ and\ \citenamefont {Wrachtrup}}]{kolesov_optical_2012}%
  \BibitemOpen
  \bibfield  {author} {\bibinfo {author} {\bibfnamefont {R.}~\bibnamefont
  {Kolesov}}, \bibinfo {author} {\bibfnamefont {K.}~\bibnamefont {Xia}},
  \bibinfo {author} {\bibfnamefont {R.}~\bibnamefont {Reuter}}, \bibinfo
  {author} {\bibfnamefont {R.}~\bibnamefont {St{\"o}hr}}, \bibinfo {author}
  {\bibfnamefont {A.}~\bibnamefont {Zappe}}, \bibinfo {author} {\bibfnamefont
  {J.}~\bibnamefont {Meijer}}, \bibinfo {author} {\bibfnamefont {P.~R.}\
  \bibnamefont {Hemmer}},\ and\ \bibinfo {author} {\bibfnamefont
  {J.}~\bibnamefont {Wrachtrup}},\ }\bibfield  {title} {\bibinfo {title}
  {Optical detection of a single rare-earth ion in a crystal},\ }\href
  {https://doi.org/10.1038/ncomms2034} {\bibfield  {journal} {\bibinfo
  {journal} {Nature Communications}\ }\textbf {\bibinfo {volume} {3}},\
  \bibinfo {pages} {1029} (\bibinfo {year} {2012})}\BibitemShut {NoStop}%
\bibitem [{\citenamefont {Kindem}\ \emph {et~al.}(2020)\citenamefont {Kindem},
  \citenamefont {Ruskuc}, \citenamefont {Bartholomew}, \citenamefont {Rochman},
  \citenamefont {Huan},\ and\ \citenamefont {Faraon}}]{kindem_control_2020}%
  \BibitemOpen
  \bibfield  {author} {\bibinfo {author} {\bibfnamefont {J.~M.}\ \bibnamefont
  {Kindem}}, \bibinfo {author} {\bibfnamefont {A.}~\bibnamefont {Ruskuc}},
  \bibinfo {author} {\bibfnamefont {J.~G.}\ \bibnamefont {Bartholomew}},
  \bibinfo {author} {\bibfnamefont {J.}~\bibnamefont {Rochman}}, \bibinfo
  {author} {\bibfnamefont {Y.~Q.}\ \bibnamefont {Huan}},\ and\ \bibinfo
  {author} {\bibfnamefont {A.}~\bibnamefont {Faraon}},\ }\bibfield  {title}
  {\bibinfo {title} {Control and single-shot readout of an ion embedded in a
  nanophotonic cavity},\ }\href {https://doi.org/10.1038/s41586-020-2160-9}
  {\bibfield  {journal} {\bibinfo  {journal} {Nature}\ }\textbf {\bibinfo
  {volume} {580}},\ \bibinfo {pages} {201} (\bibinfo {year}
  {2020})}\BibitemShut {NoStop}%
\bibitem [{\citenamefont {Jobez}\ \emph {et~al.}(2014)\citenamefont {Jobez},
  \citenamefont {Usmani}, \citenamefont {Timoney}, \citenamefont {Laplane},
  \citenamefont {Gisin},\ and\ \citenamefont
  {Afzelius}}]{jobez_cavity-enhanced_2014}%
  \BibitemOpen
  \bibfield  {author} {\bibinfo {author} {\bibfnamefont {P.}~\bibnamefont
  {Jobez}}, \bibinfo {author} {\bibfnamefont {I.}~\bibnamefont {Usmani}},
  \bibinfo {author} {\bibfnamefont {N.}~\bibnamefont {Timoney}}, \bibinfo
  {author} {\bibfnamefont {C.}~\bibnamefont {Laplane}}, \bibinfo {author}
  {\bibfnamefont {N.}~\bibnamefont {Gisin}},\ and\ \bibinfo {author}
  {\bibfnamefont {M.}~\bibnamefont {Afzelius}},\ }\bibfield  {title} {\bibinfo
  {title} {Cavity-enhanced storage in an optical spin-wave memory},\ }\href
  {https://doi.org/10.1088/1367-2630/16/8/083005} {\bibfield  {journal}
  {\bibinfo  {journal} {New Journal of Physics}\ }\textbf {\bibinfo {volume}
  {16}},\ \bibinfo {pages} {083005} (\bibinfo {year} {2014})}\BibitemShut
  {NoStop}%
\bibitem [{\citenamefont {Williamson}\ \emph {et~al.}(2014)\citenamefont
  {Williamson}, \citenamefont {Chen},\ and\ \citenamefont
  {Longdell}}]{williamson_magneto-optic_2014}%
  \BibitemOpen
  \bibfield  {author} {\bibinfo {author} {\bibfnamefont {L.~A.}\ \bibnamefont
  {Williamson}}, \bibinfo {author} {\bibfnamefont {Y.-H.}\ \bibnamefont
  {Chen}},\ and\ \bibinfo {author} {\bibfnamefont {J.~J.}\ \bibnamefont
  {Longdell}},\ }\bibfield  {title} {\bibinfo {title} {Magneto-{{Optic
  Modulator}} with {{Unit Quantum Efficiency}}},\ }\href
  {https://doi.org/10.1103/PhysRevLett.113.203601} {\bibfield  {journal}
  {\bibinfo  {journal} {Physical Review Letters}\ }\textbf {\bibinfo {volume}
  {113}},\ \bibinfo {pages} {203601} (\bibinfo {year} {2014})}\BibitemShut
  {NoStop}%
\bibitem [{\citenamefont {Miyazono}\ \emph {et~al.}(2016)\citenamefont
  {Miyazono}, \citenamefont {Zhong}, \citenamefont {Craiciu}, \citenamefont
  {Kindem},\ and\ \citenamefont {Faraon}}]{miyazono_coupling_2016}%
  \BibitemOpen
  \bibfield  {author} {\bibinfo {author} {\bibfnamefont {E.}~\bibnamefont
  {Miyazono}}, \bibinfo {author} {\bibfnamefont {T.}~\bibnamefont {Zhong}},
  \bibinfo {author} {\bibfnamefont {I.}~\bibnamefont {Craiciu}}, \bibinfo
  {author} {\bibfnamefont {J.~M.}\ \bibnamefont {Kindem}},\ and\ \bibinfo
  {author} {\bibfnamefont {A.}~\bibnamefont {Faraon}},\ }\bibfield  {title}
  {\bibinfo {title} {Coupling of erbium dopants to yttrium orthosilicate
  photonic crystal cavities for on-chip optical quantum memories},\ }\href
  {https://doi.org/10.1063/1.4939651} {\bibfield  {journal} {\bibinfo
  {journal} {Applied Physics Letters}\ }\textbf {\bibinfo {volume} {108}},\
  \bibinfo {pages} {011111} (\bibinfo {year} {2016})}\BibitemShut {NoStop}%
\bibitem [{\citenamefont {Fernandez-Gonzalvo}\ \emph
  {et~al.}(2019)\citenamefont {Fernandez-Gonzalvo}, \citenamefont {Horvath},
  \citenamefont {Chen},\ and\ \citenamefont
  {Longdell}}]{fernandez-gonzalvo_cavity-enhanced_2019}%
  \BibitemOpen
  \bibfield  {author} {\bibinfo {author} {\bibfnamefont {X.}~\bibnamefont
  {Fernandez-Gonzalvo}}, \bibinfo {author} {\bibfnamefont {S.~P.}\ \bibnamefont
  {Horvath}}, \bibinfo {author} {\bibfnamefont {Y.-H.}\ \bibnamefont {Chen}},\
  and\ \bibinfo {author} {\bibfnamefont {J.~J.}\ \bibnamefont {Longdell}},\
  }\bibfield  {title} {\bibinfo {title} {Cavity-enhanced raman heterodyne
  spectroscopy in {Er$^{3+}$:Y$_2$SiO$_5$} for microwave to optical signal
  conversion},\ }\href {https://doi.org/10.1103/PhysRevA.100.033807} {\bibfield
   {journal} {\bibinfo  {journal} {Physical Review A}\ }\textbf {\bibinfo
  {volume} {100}},\ \bibinfo {pages} {033807} (\bibinfo {year}
  {2019})}\BibitemShut {NoStop}%
\bibitem [{\citenamefont {Strekalov}\ \emph {et~al.}(2016)\citenamefont
  {Strekalov}, \citenamefont {Marquardt}, \citenamefont {Matsko}, \citenamefont
  {Schwefel},\ and\ \citenamefont {Leuchs}}]{strekalov_nonlinear_2016}%
  \BibitemOpen
  \bibfield  {author} {\bibinfo {author} {\bibfnamefont {D.~V.}\ \bibnamefont
  {Strekalov}}, \bibinfo {author} {\bibfnamefont {C.}~\bibnamefont
  {Marquardt}}, \bibinfo {author} {\bibfnamefont {A.~B.}\ \bibnamefont
  {Matsko}}, \bibinfo {author} {\bibfnamefont {H.~G.~L.}\ \bibnamefont
  {Schwefel}},\ and\ \bibinfo {author} {\bibfnamefont {G.}~\bibnamefont
  {Leuchs}},\ }\bibfield  {title} {\bibinfo {title} {Nonlinear and quantum
  optics with whispering gallery resonators},\ }\href
  {https://doi.org/10.1088/2040-8978/18/12/123002} {\bibfield  {journal}
  {\bibinfo  {journal} {Journal of Optics}\ }\textbf {\bibinfo {volume} {18}},\
  \bibinfo {pages} {123002} (\bibinfo {year} {2016})}\BibitemShut {NoStop}%
\bibitem [{\citenamefont {Norman}\ \emph {et~al.}(2022)\citenamefont {Norman},
  \citenamefont {Azeem}, \citenamefont {Longdell},\ and\ \citenamefont
  {Schwefel}}]{norman_measuring_2022}%
  \BibitemOpen
  \bibfield  {author} {\bibinfo {author} {\bibfnamefont {D.~S.}\ \bibnamefont
  {Norman}}, \bibinfo {author} {\bibfnamefont {F.}~\bibnamefont {Azeem}},
  \bibinfo {author} {\bibfnamefont {J.~J.}\ \bibnamefont {Longdell}},\ and\
  \bibinfo {author} {\bibfnamefont {H.~G.~L.}\ \bibnamefont {Schwefel}},\
  }\bibfield  {title} {\bibinfo {title} {Measuring optical loss in yttrium
  orthosilicate using a whispering gallery mode resonator},\ }\href
  {https://doi.org/10.1088/2633-4356/ac4c39} {\bibfield  {journal} {\bibinfo
  {journal} {Materials for Quantum Technology}\ }\textbf {\bibinfo {volume}
  {2}},\ \bibinfo {pages} {011001} (\bibinfo {year} {2022})}\BibitemShut
  {NoStop}%
\bibitem [{\citenamefont {Kuo}\ \emph {et~al.}(2014)\citenamefont {Kuo},
  \citenamefont {Bravo-Abad},\ and\ \citenamefont
  {Solomon}}]{kuo_second-harmonic_2014}%
  \BibitemOpen
  \bibfield  {author} {\bibinfo {author} {\bibfnamefont {P.~S.}\ \bibnamefont
  {Kuo}}, \bibinfo {author} {\bibfnamefont {J.}~\bibnamefont {Bravo-Abad}},\
  and\ \bibinfo {author} {\bibfnamefont {G.~S.}\ \bibnamefont {Solomon}},\
  }\bibfield  {title} {\bibinfo {title} {Second-harmonic generation using
  {$\bar{4}$}-quasi-phasematching in a {GaAs} whispering-gallery-mode
  microcavity},\ }\href {https://doi.org/10.1038/ncomms4109} {\bibfield
  {journal} {\bibinfo  {journal} {Nature Communications}\ }\textbf {\bibinfo
  {volume} {5}},\ \bibinfo {pages} {3109} (\bibinfo {year} {2014})}\BibitemShut
  {NoStop}%
\bibitem [{\citenamefont {F{\"u}rst}\ \emph {et~al.}(2015)\citenamefont
  {F{\"u}rst}, \citenamefont {Buse}, \citenamefont {Breunig}, \citenamefont
  {Becker}, \citenamefont {Liebertz},\ and\ \citenamefont
  {Bohat{\'y}}}]{furst_second-harmonic_2015}%
  \BibitemOpen
  \bibfield  {author} {\bibinfo {author} {\bibfnamefont {J.~U.}\ \bibnamefont
  {F{\"u}rst}}, \bibinfo {author} {\bibfnamefont {K.}~\bibnamefont {Buse}},
  \bibinfo {author} {\bibfnamefont {I.}~\bibnamefont {Breunig}}, \bibinfo
  {author} {\bibfnamefont {P.}~\bibnamefont {Becker}}, \bibinfo {author}
  {\bibfnamefont {J.}~\bibnamefont {Liebertz}},\ and\ \bibinfo {author}
  {\bibfnamefont {L.}~\bibnamefont {Bohat{\'y}}},\ }\bibfield  {title}
  {\bibinfo {title} {Second-harmonic generation of light at 245 nm in a lithium
  tetraborate whispering gallery resonator},\ }\href
  {https://doi.org/10.1364/OL.40.001932} {\bibfield  {journal} {\bibinfo
  {journal} {Optics Letters}\ }\textbf {\bibinfo {volume} {40}},\ \bibinfo
  {pages} {1932} (\bibinfo {year} {2015})}\BibitemShut {NoStop}%
\bibitem [{\citenamefont {Trainor}\ \emph {et~al.}(2018)\citenamefont
  {Trainor}, \citenamefont {Sedlmeir}, \citenamefont {Peuntinger},\ and\
  \citenamefont {Schwefel}}]{trainor_selective_2018}%
  \BibitemOpen
  \bibfield  {author} {\bibinfo {author} {\bibfnamefont {L.~S.}\ \bibnamefont
  {Trainor}}, \bibinfo {author} {\bibfnamefont {F.}~\bibnamefont {Sedlmeir}},
  \bibinfo {author} {\bibfnamefont {C.}~\bibnamefont {Peuntinger}},\ and\
  \bibinfo {author} {\bibfnamefont {H.~G.~L.}\ \bibnamefont {Schwefel}},\
  }\bibfield  {title} {\bibinfo {title} {Selective {{Coupling Enhances Harmonic
  Generation}} of {{Whispering-Gallery Modes}}},\ }\href
  {https://doi.org/10.1103/PhysRevApplied.9.024007} {\bibfield  {journal}
  {\bibinfo  {journal} {Physical Review Applied}\ }\textbf {\bibinfo {volume}
  {9}},\ \bibinfo {pages} {024007} (\bibinfo {year} {2018})}\BibitemShut
  {NoStop}%
\bibitem [{\citenamefont {Herr}\ \emph {et~al.}(2014)\citenamefont {Herr},
  \citenamefont {Brasch}, \citenamefont {Jost}, \citenamefont {Wang},
  \citenamefont {Kondratiev}, \citenamefont {Gorodetsky},\ and\ \citenamefont
  {Kippenberg}}]{herr_temporal_2014}%
  \BibitemOpen
  \bibfield  {author} {\bibinfo {author} {\bibfnamefont {T.}~\bibnamefont
  {Herr}}, \bibinfo {author} {\bibfnamefont {V.}~\bibnamefont {Brasch}},
  \bibinfo {author} {\bibfnamefont {J.~D.}\ \bibnamefont {Jost}}, \bibinfo
  {author} {\bibfnamefont {C.~Y.}\ \bibnamefont {Wang}}, \bibinfo {author}
  {\bibfnamefont {N.~M.}\ \bibnamefont {Kondratiev}}, \bibinfo {author}
  {\bibfnamefont {M.~L.}\ \bibnamefont {Gorodetsky}},\ and\ \bibinfo {author}
  {\bibfnamefont {T.~J.}\ \bibnamefont {Kippenberg}},\ }\bibfield  {title}
  {\bibinfo {title} {Temporal solitons in optical microresonators},\ }\href
  {https://doi.org/10.1038/nphoton.2013.343} {\bibfield  {journal} {\bibinfo
  {journal} {Nature Photonics}\ }\textbf {\bibinfo {volume} {8}},\ \bibinfo
  {pages} {145} (\bibinfo {year} {2014})}\BibitemShut {NoStop}%
\bibitem [{\citenamefont {Webb}\ \emph {et~al.}(2016)\citenamefont {Webb},
  \citenamefont {Erkintalo}, \citenamefont {Coen},\ and\ \citenamefont
  {Murdoch}}]{webb_experimental_2016}%
  \BibitemOpen
  \bibfield  {author} {\bibinfo {author} {\bibfnamefont {K.~E.}\ \bibnamefont
  {Webb}}, \bibinfo {author} {\bibfnamefont {M.}~\bibnamefont {Erkintalo}},
  \bibinfo {author} {\bibfnamefont {S.}~\bibnamefont {Coen}},\ and\ \bibinfo
  {author} {\bibfnamefont {S.~G.}\ \bibnamefont {Murdoch}},\ }\bibfield
  {title} {\bibinfo {title} {Experimental observation of coherent cavity
  soliton frequency combs in silica microspheres},\ }\href
  {https://doi.org/10.1364/OL.41.004613} {\bibfield  {journal} {\bibinfo
  {journal} {Optics Letters}\ }\textbf {\bibinfo {volume} {41}},\ \bibinfo
  {pages} {4613} (\bibinfo {year} {2016})}\BibitemShut {NoStop}%
\bibitem [{\citenamefont {Meisenheimer}\ \emph {et~al.}(2017)\citenamefont
  {Meisenheimer}, \citenamefont {Fürst}, \citenamefont {Buse},\ and\
  \citenamefont {Breunig}}]{meisenheimer_continuous-wave_2017}%
  \BibitemOpen
  \bibfield  {author} {\bibinfo {author} {\bibfnamefont {S.-K.}\ \bibnamefont
  {Meisenheimer}}, \bibinfo {author} {\bibfnamefont {J.~U.}\ \bibnamefont
  {F{\"u}rst}}, \bibinfo {author} {\bibfnamefont {K.}~\bibnamefont {Buse}},\ and\
  \bibinfo {author} {\bibfnamefont {I.}~\bibnamefont {Breunig}},\ }\bibfield
  {title} {\bibinfo {title} {Continuous-wave optical parametric oscillation
  tunable up to an 8 {$\mu$}m wavelength},\ }\href
  {https://doi.org/10.1364/OPTICA.4.000189} {\bibfield  {journal} {\bibinfo
  {journal} {Optica}\ }\textbf {\bibinfo {volume} {4}},\ \bibinfo {pages} {189}
  (\bibinfo {year} {2017})}\BibitemShut {NoStop}%
\bibitem [{\citenamefont {F{\"o}rtsch}\ \emph {et~al.}(2015)\citenamefont
  {F{\"o}rtsch}, \citenamefont {Schunk}, \citenamefont {F{\"u}rst},
  \citenamefont {Strekalov}, \citenamefont {Gerrits}, \citenamefont {Stevens},
  \citenamefont {Sedlmeir}, \citenamefont {Schwefel}, \citenamefont {Nam},
  \citenamefont {Leuchs},\ and\ \citenamefont
  {Marquardt}}]{fortsch_highly_2015}%
  \BibitemOpen
  \bibfield  {author} {\bibinfo {author} {\bibfnamefont {M.}~\bibnamefont
  {F{\"o}rtsch}}, \bibinfo {author} {\bibfnamefont {G.}~\bibnamefont {Schunk}},
  \bibinfo {author} {\bibfnamefont {J.~U.}\ \bibnamefont {F{\"u}rst}}, \bibinfo
  {author} {\bibfnamefont {D.}~\bibnamefont {Strekalov}}, \bibinfo {author}
  {\bibfnamefont {T.}~\bibnamefont {Gerrits}}, \bibinfo {author} {\bibfnamefont
  {M.~J.}\ \bibnamefont {Stevens}}, \bibinfo {author} {\bibfnamefont
  {F.}~\bibnamefont {Sedlmeir}}, \bibinfo {author} {\bibfnamefont {H.~G.~L.}\
  \bibnamefont {Schwefel}}, \bibinfo {author} {\bibfnamefont {S.~W.}\
  \bibnamefont {Nam}}, \bibinfo {author} {\bibfnamefont {G.}~\bibnamefont
  {Leuchs}},\ and\ \bibinfo {author} {\bibfnamefont {C.}~\bibnamefont
  {Marquardt}},\ }\bibfield  {title} {\bibinfo {title} {Highly efficient
  generation of single-mode photon pairs from a crystalline
  whispering-gallery-mode resonator source},\ }\href
  {https://doi.org/10.1103/PhysRevA.91.023812} {\bibfield  {journal} {\bibinfo
  {journal} {Physical Review A}\ }\textbf {\bibinfo {volume} {91}},\ \bibinfo
  {pages} {023812} (\bibinfo {year} {2015})}\BibitemShut {NoStop}%
\bibitem [{\citenamefont {F{\"u}rst}\ \emph {et~al.}(2011)\citenamefont
  {F{\"u}rst}, \citenamefont {Strekalov}, \citenamefont {Elser}, \citenamefont
  {Aiello}, \citenamefont {Andersen}, \citenamefont {Marquardt},\ and\
  \citenamefont {Leuchs}}]{furst_quantum_2011}%
  \BibitemOpen
  \bibfield  {author} {\bibinfo {author} {\bibfnamefont {J.~U.}\ \bibnamefont
  {F{\"u}rst}}, \bibinfo {author} {\bibfnamefont {D.~V.}\ \bibnamefont
  {Strekalov}}, \bibinfo {author} {\bibfnamefont {D.}~\bibnamefont {Elser}},
  \bibinfo {author} {\bibfnamefont {A.}~\bibnamefont {Aiello}}, \bibinfo
  {author} {\bibfnamefont {U.~L.}\ \bibnamefont {Andersen}}, \bibinfo {author}
  {\bibfnamefont {C.}~\bibnamefont {Marquardt}},\ and\ \bibinfo {author}
  {\bibfnamefont {G.}~\bibnamefont {Leuchs}},\ }\bibfield  {title} {\bibinfo
  {title} {Quantum {{Light}} from a {{Whispering-Gallery-Mode Disk
  Resonator}}},\ }\href {https://doi.org/10.1103/PhysRevLett.106.113901}
  {\bibfield  {journal} {\bibinfo  {journal} {Physical Review Letters}\
  }\textbf {\bibinfo {volume} {106}},\ \bibinfo {pages} {113901} (\bibinfo
  {year} {2011})}\BibitemShut {NoStop}%
\bibitem [{\citenamefont {Savchenkov}\ \emph {et~al.}(2009)\citenamefont
  {Savchenkov}, \citenamefont {Liang}, \citenamefont {Matsko}, \citenamefont
  {Ilchenko}, \citenamefont {Seidel},\ and\ \citenamefont
  {Maleki}}]{savchenkov_tunable_2009}%
  \BibitemOpen
  \bibfield  {author} {\bibinfo {author} {\bibfnamefont {A.~A.}\ \bibnamefont
  {Savchenkov}}, \bibinfo {author} {\bibfnamefont {W.}~\bibnamefont {Liang}},
  \bibinfo {author} {\bibfnamefont {A.~B.}\ \bibnamefont {Matsko}}, \bibinfo
  {author} {\bibfnamefont {V.~S.}\ \bibnamefont {Ilchenko}}, \bibinfo {author}
  {\bibfnamefont {D.}~\bibnamefont {Seidel}},\ and\ \bibinfo {author}
  {\bibfnamefont {L.}~\bibnamefont {Maleki}},\ }\bibfield  {title} {\bibinfo
  {title} {Tunable optical single-sideband modulator with complete sideband
  suppression},\ }\href {https://doi.org/10.1364/OL.34.001300} {\bibfield
  {journal} {\bibinfo  {journal} {Optics Letters}\ }\textbf {\bibinfo {volume}
  {34}},\ \bibinfo {pages} {1300} (\bibinfo {year} {2009})}\BibitemShut
  {NoStop}%
\bibitem [{\citenamefont {Zhang}\ \emph {et~al.}(2019)\citenamefont {Zhang},
  \citenamefont {Buscaino}, \citenamefont {Wang}, \citenamefont
  {{Shams-Ansari}}, \citenamefont {Reimer}, \citenamefont {Zhu}, \citenamefont
  {Kahn},\ and\ \citenamefont {Lon{\v c}ar}}]{zhang_broadband_2019}%
  \BibitemOpen
  \bibfield  {author} {\bibinfo {author} {\bibfnamefont {M.}~\bibnamefont
  {Zhang}}, \bibinfo {author} {\bibfnamefont {B.}~\bibnamefont {Buscaino}},
  \bibinfo {author} {\bibfnamefont {C.}~\bibnamefont {Wang}}, \bibinfo {author}
  {\bibfnamefont {A.}~\bibnamefont {{Shams-Ansari}}}, \bibinfo {author}
  {\bibfnamefont {C.}~\bibnamefont {Reimer}}, \bibinfo {author} {\bibfnamefont
  {R.}~\bibnamefont {Zhu}}, \bibinfo {author} {\bibfnamefont {J.~M.}\
  \bibnamefont {Kahn}},\ and\ \bibinfo {author} {\bibfnamefont
  {M.}~\bibnamefont {Lon{\v c}ar}},\ }\bibfield  {title} {\bibinfo {title}
  {Broadband electro-optic frequency comb generation in a lithium niobate
  microring resonator},\ }\href {https://doi.org/10.1038/s41586-019-1008-7}
  {\bibfield  {journal} {\bibinfo  {journal} {Nature}\ }\textbf {\bibinfo
  {volume} {568}},\ \bibinfo {pages} {373} (\bibinfo {year}
  {2019})}\BibitemShut {NoStop}%
\bibitem [{\citenamefont {Rueda}\ \emph {et~al.}(2019)\citenamefont {Rueda},
  \citenamefont {Sedlmeir}, \citenamefont {Kumari}, \citenamefont {Leuchs},\
  and\ \citenamefont {Schwefel}}]{rueda_resonant_2019}%
  \BibitemOpen
  \bibfield  {author} {\bibinfo {author} {\bibfnamefont {A.}~\bibnamefont
  {Rueda}}, \bibinfo {author} {\bibfnamefont {F.}~\bibnamefont {Sedlmeir}},
  \bibinfo {author} {\bibfnamefont {M.}~\bibnamefont {Kumari}}, \bibinfo
  {author} {\bibfnamefont {G.}~\bibnamefont {Leuchs}},\ and\ \bibinfo {author}
  {\bibfnamefont {H.~G.~L.}\ \bibnamefont {Schwefel}},\ }\bibfield  {title}
  {\bibinfo {title} {Resonant electro-optic frequency comb},\ }\href
  {https://doi.org/10.1038/s41586-019-1110-x} {\bibfield  {journal} {\bibinfo
  {journal} {Nature}\ }\textbf {\bibinfo {volume} {568}},\ \bibinfo {pages}
  {378} (\bibinfo {year} {2019})}\BibitemShut {NoStop}%
\bibitem [{\citenamefont {Haigh}\ \emph {et~al.}(2018)\citenamefont {Haigh},
  \citenamefont {Lambert}, \citenamefont {Sharma}, \citenamefont {Blanter},
  \citenamefont {Bauer},\ and\ \citenamefont {Ramsay}}]{haigh_selection_2018}%
  \BibitemOpen
  \bibfield  {author} {\bibinfo {author} {\bibfnamefont {J.~A.}\ \bibnamefont
  {Haigh}}, \bibinfo {author} {\bibfnamefont {N.~J.}\ \bibnamefont {Lambert}},
  \bibinfo {author} {\bibfnamefont {S.}~\bibnamefont {Sharma}}, \bibinfo
  {author} {\bibfnamefont {Y.~M.}\ \bibnamefont {Blanter}}, \bibinfo {author}
  {\bibfnamefont {G.~E.~W.}\ \bibnamefont {Bauer}},\ and\ \bibinfo {author}
  {\bibfnamefont {A.~J.}\ \bibnamefont {Ramsay}},\ }\bibfield  {title}
  {\bibinfo {title} {Selection rules for cavity-enhanced {{Brillouin}} light
  scattering from magnetostatic modes},\ }\href
  {https://doi.org/10.1103/PhysRevB.97.214423} {\bibfield  {journal} {\bibinfo
  {journal} {Physical Review B}\ }\textbf {\bibinfo {volume} {97}},\ \bibinfo
  {pages} {214423} (\bibinfo {year} {2018})}\BibitemShut {NoStop}%
\bibitem [{\citenamefont {McAuslan}\ \emph {et~al.}(2011)\citenamefont
  {McAuslan}, \citenamefont {Korystov},\ and\ \citenamefont
  {Longdell}}]{mcauslan_coherent_2011}%
  \BibitemOpen
  \bibfield  {author} {\bibinfo {author} {\bibfnamefont {D.~L.}\ \bibnamefont
  {McAuslan}}, \bibinfo {author} {\bibfnamefont {D.}~\bibnamefont {Korystov}},\
  and\ \bibinfo {author} {\bibfnamefont {J.~J.}\ \bibnamefont {Longdell}},\
  }\bibfield  {title} {\bibinfo {title} {Coherent spectroscopy of
  rare-earth-metal-ion-doped whispering-gallery-mode resonators},\ }\href
  {https://doi.org/10.1103/PhysRevA.83.063847} {\bibfield  {journal} {\bibinfo
  {journal} {Physical Review A}\ }\textbf {\bibinfo {volume} {83}},\ \bibinfo
  {pages} {063847} (\bibinfo {year} {2011})}\BibitemShut {NoStop}%
\bibitem [{\citenamefont {Carbonera}(2009)}]{carbonera_optically_2009}%
  \BibitemOpen
  \bibfield  {author} {\bibinfo {author} {\bibfnamefont {D.}~\bibnamefont
  {Carbonera}},\ }\bibfield  {title} {\bibinfo {title} {Optically detected
  magnetic resonance ({{ODMR}}) of photoexcited triplet states},\ }\href
  {https://doi.org/10.1007/s11120-009-9407-5} {\bibfield  {journal} {\bibinfo
  {journal} {Photosynthesis Research}\ }\textbf {\bibinfo {volume} {102}},\
  \bibinfo {pages} {403} (\bibinfo {year} {2009})}\BibitemShut {NoStop}%
\bibitem [{\citenamefont {Suter}(2020)}]{suter_optical_2020}%
  \BibitemOpen
  \bibfield  {author} {\bibinfo {author} {\bibfnamefont {D.}~\bibnamefont
  {Suter}},\ }\bibfield  {title} {\bibinfo {title} {Optical detection of
  magnetic resonance},\ }\href {https://doi.org/10.5194/mr-1-115-2020}
  {\bibfield  {journal} {\bibinfo  {journal} {Magnetic Resonance}\ }\textbf
  {\bibinfo {volume} {1}},\ \bibinfo {pages} {115} (\bibinfo {year}
  {2020})}\BibitemShut {NoStop}%
\bibitem [{\citenamefont {Chernobrod}\ and\ \citenamefont
  {Berman}(2005)}]{chernobrod_spin_2005}%
  \BibitemOpen
  \bibfield  {author} {\bibinfo {author} {\bibfnamefont {B.~M.}\ \bibnamefont
  {Chernobrod}}\ and\ \bibinfo {author} {\bibfnamefont {G.~P.}\ \bibnamefont
  {Berman}},\ }\bibfield  {title} {\bibinfo {title} {Spin microscope based on
  optically detected magnetic resonance},\ }\href
  {https://doi.org/10.1063/1.1829373} {\bibfield  {journal} {\bibinfo
  {journal} {Journal of Applied Physics}\ }\textbf {\bibinfo {volume} {97}},\
  \bibinfo {pages} {014903} (\bibinfo {year} {2005})}\BibitemShut {NoStop}%
\bibitem [{\citenamefont {Jacques}\ \emph {et~al.}(2009)\citenamefont
  {Jacques}, \citenamefont {Neumann}, \citenamefont {Beck}, \citenamefont
  {Markham}, \citenamefont {Twitchen}, \citenamefont {Meijer}, \citenamefont
  {Kaiser}, \citenamefont {Balasubramanian}, \citenamefont {Jelezko},\ and\
  \citenamefont {Wrachtrup}}]{jacques_dynamic_2009}%
  \BibitemOpen
  \bibfield  {author} {\bibinfo {author} {\bibfnamefont {V.}~\bibnamefont
  {Jacques}}, \bibinfo {author} {\bibfnamefont {P.}~\bibnamefont {Neumann}},
  \bibinfo {author} {\bibfnamefont {J.}~\bibnamefont {Beck}}, \bibinfo {author}
  {\bibfnamefont {M.}~\bibnamefont {Markham}}, \bibinfo {author} {\bibfnamefont
  {D.}~\bibnamefont {Twitchen}}, \bibinfo {author} {\bibfnamefont
  {J.}~\bibnamefont {Meijer}}, \bibinfo {author} {\bibfnamefont
  {F.}~\bibnamefont {Kaiser}}, \bibinfo {author} {\bibfnamefont
  {G.}~\bibnamefont {Balasubramanian}}, \bibinfo {author} {\bibfnamefont
  {F.}~\bibnamefont {Jelezko}},\ and\ \bibinfo {author} {\bibfnamefont
  {J.}~\bibnamefont {Wrachtrup}},\ }\bibfield  {title} {\bibinfo {title}
  {Dynamic {{Polarization}} of {{Single Nuclear Spins}} by {{Optical Pumping}}
  of {{Nitrogen-Vacancy Color Centers}} in {{Diamond}} at {{Room
  Temperature}}},\ }\href {https://doi.org/10.1103/PhysRevLett.102.057403}
  {\bibfield  {journal} {\bibinfo  {journal} {Physical Review Letters}\
  }\textbf {\bibinfo {volume} {102}},\ \bibinfo {pages} {057403} (\bibinfo
  {year} {2009})}\BibitemShut {NoStop}%
\bibitem [{\citenamefont {Siyushev}\ \emph {et~al.}(2014)\citenamefont
  {Siyushev}, \citenamefont {Xia}, \citenamefont {Reuter}, \citenamefont
  {Jamali}, \citenamefont {Zhao}, \citenamefont {Yang}, \citenamefont {Duan},
  \citenamefont {Kukharchyk}, \citenamefont {Wieck}, \citenamefont {Kolesov},\
  and\ \citenamefont {Wrachtrup}}]{siyushev_coherent_2014}%
  \BibitemOpen
  \bibfield  {author} {\bibinfo {author} {\bibfnamefont {P.}~\bibnamefont
  {Siyushev}}, \bibinfo {author} {\bibfnamefont {K.}~\bibnamefont {Xia}},
  \bibinfo {author} {\bibfnamefont {R.}~\bibnamefont {Reuter}}, \bibinfo
  {author} {\bibfnamefont {M.}~\bibnamefont {Jamali}}, \bibinfo {author}
  {\bibfnamefont {N.}~\bibnamefont {Zhao}}, \bibinfo {author} {\bibfnamefont
  {N.}~\bibnamefont {Yang}}, \bibinfo {author} {\bibfnamefont {C.}~\bibnamefont
  {Duan}}, \bibinfo {author} {\bibfnamefont {N.}~\bibnamefont {Kukharchyk}},
  \bibinfo {author} {\bibfnamefont {A.~D.}\ \bibnamefont {Wieck}}, \bibinfo
  {author} {\bibfnamefont {R.}~\bibnamefont {Kolesov}},\ and\ \bibinfo {author}
  {\bibfnamefont {J.}~\bibnamefont {Wrachtrup}},\ }\bibfield  {title} {\bibinfo
  {title} {Coherent properties of single rare-earth spin qubits},\ }\href
  {https://doi.org/10.1038/ncomms4895} {\bibfield  {journal} {\bibinfo
  {journal} {Nature Communications}\ }\textbf {\bibinfo {volume} {5}},\
  \bibinfo {pages} {3895} (\bibinfo {year} {2014})}\BibitemShut {NoStop}%
\bibitem [{\citenamefont {Lambert}\ \emph {et~al.}(2020)\citenamefont
  {Lambert}, \citenamefont {Rueda}, \citenamefont {Sedlmeir},\ and\
  \citenamefont {Schwefel}}]{lambert_coherent_2020}%
  \BibitemOpen
  \bibfield  {author} {\bibinfo {author} {\bibfnamefont {N.~J.}\ \bibnamefont
  {Lambert}}, \bibinfo {author} {\bibfnamefont {A.}~\bibnamefont {Rueda}},
  \bibinfo {author} {\bibfnamefont {F.}~\bibnamefont {Sedlmeir}},\ and\
  \bibinfo {author} {\bibfnamefont {H.~G.~L.}\ \bibnamefont {Schwefel}},\
  }\bibfield  {title} {\bibinfo {title} {Coherent {{Conversion Between
  Microwave}} and {{Optical Photons}}\textemdash{{An Overview}} of {{Physical
  Implementations}}},\ }\href {https://doi.org/10.1002/qute.201900077}
  {\bibfield  {journal} {\bibinfo  {journal} {Advanced Quantum Technologies}\
  }\textbf {\bibinfo {volume} {3}},\ \bibinfo {pages} {1900077} (\bibinfo
  {year} {2020})}\BibitemShut {NoStop}%
\bibitem [{\citenamefont {Lauk}\ \emph {et~al.}(2020)\citenamefont {Lauk},
  \citenamefont {Sinclair}, \citenamefont {Barzanjeh}, \citenamefont {Covey},
  \citenamefont {Saffman}, \citenamefont {Spiropulu},\ and\ \citenamefont
  {Simon}}]{lauk_perspectives_2020}%
  \BibitemOpen
  \bibfield  {author} {\bibinfo {author} {\bibfnamefont {N.}~\bibnamefont
  {Lauk}}, \bibinfo {author} {\bibfnamefont {N.}~\bibnamefont {Sinclair}},
  \bibinfo {author} {\bibfnamefont {S.}~\bibnamefont {Barzanjeh}}, \bibinfo
  {author} {\bibfnamefont {J.~P.}\ \bibnamefont {Covey}}, \bibinfo {author}
  {\bibfnamefont {M.}~\bibnamefont {Saffman}}, \bibinfo {author} {\bibfnamefont
  {M.}~\bibnamefont {Spiropulu}},\ and\ \bibinfo {author} {\bibfnamefont
  {C.}~\bibnamefont {Simon}},\ }\bibfield  {title} {\bibinfo {title}
  {Perspectives on quantum transduction},\ }\href
  {https://doi.org/10.1088/2058-9565/ab788a} {\bibfield  {journal} {\bibinfo
  {journal} {Quantum Science and Technology}\ }\textbf {\bibinfo {volume}
  {5}},\ \bibinfo {pages} {020501} (\bibinfo {year} {2020})}\BibitemShut
  {NoStop}%
\bibitem [{\citenamefont {Magnard}\ \emph {et~al.}(2020)\citenamefont
  {Magnard}, \citenamefont {Storz}, \citenamefont {Kurpiers}, \citenamefont
  {Sch{\"a}r}, \citenamefont {Marxer}, \citenamefont {L{\"u}tolf},
  \citenamefont {Walter}, \citenamefont {Besse}, \citenamefont {Gabureac},
  \citenamefont {Reuer}, \citenamefont {Akin}, \citenamefont {Royer},
  \citenamefont {Blais},\ and\ \citenamefont
  {Wallraff}}]{magnard_microwave_2020}%
  \BibitemOpen
  \bibfield  {author} {\bibinfo {author} {\bibfnamefont {P.}~\bibnamefont
  {Magnard}}, \bibinfo {author} {\bibfnamefont {S.}~\bibnamefont {Storz}},
  \bibinfo {author} {\bibfnamefont {P.}~\bibnamefont {Kurpiers}}, \bibinfo
  {author} {\bibfnamefont {J.}~\bibnamefont {Sch{\"a}r}}, \bibinfo {author}
  {\bibfnamefont {F.}~\bibnamefont {Marxer}}, \bibinfo {author} {\bibfnamefont
  {J.}~\bibnamefont {L{\"u}tolf}}, \bibinfo {author} {\bibfnamefont
  {T.}~\bibnamefont {Walter}}, \bibinfo {author} {\bibfnamefont {J.-C.}\
  \bibnamefont {Besse}}, \bibinfo {author} {\bibfnamefont {M.}~\bibnamefont
  {Gabureac}}, \bibinfo {author} {\bibfnamefont {K.}~\bibnamefont {Reuer}},
  \bibinfo {author} {\bibfnamefont {A.}~\bibnamefont {Akin}}, \bibinfo {author}
  {\bibfnamefont {B.}~\bibnamefont {Royer}}, \bibinfo {author} {\bibfnamefont
  {A.}~\bibnamefont {Blais}},\ and\ \bibinfo {author} {\bibfnamefont
  {A.}~\bibnamefont {Wallraff}},\ }\bibfield  {title} {\bibinfo {title}
  {Microwave {{Quantum Link}} between {{Superconducting Circuits Housed}} in
  {{Spatially Separated Cryogenic Systems}}},\ }\href
  {https://doi.org/10.1103/PhysRevLett.125.260502} {\bibfield  {journal}
  {\bibinfo  {journal} {Physical Review Letters}\ }\textbf {\bibinfo {volume}
  {125}},\ \bibinfo {pages} {260502} (\bibinfo {year} {2020})}\BibitemShut
  {NoStop}%
\bibitem [{\citenamefont {Grudinin}\ \emph {et~al.}(2006)\citenamefont
  {Grudinin}, \citenamefont {Ilchenko},\ and\ \citenamefont
  {Maleki}}]{grudinin_ultrahigh_2006}%
  \BibitemOpen
  \bibfield  {author} {\bibinfo {author} {\bibfnamefont {I.~S.}\ \bibnamefont
  {Grudinin}}, \bibinfo {author} {\bibfnamefont {V.~S.}\ \bibnamefont
  {Ilchenko}},\ and\ \bibinfo {author} {\bibfnamefont {L.}~\bibnamefont
  {Maleki}},\ }\bibfield  {title} {\bibinfo {title} {Ultrahigh optical {{Q}}
  factors of crystalline resonators in the linear regime},\ }\href
  {https://doi.org/10.1103/PhysRevA.74.063806} {\bibfield  {journal} {\bibinfo
  {journal} {Physical Review A}\ }\textbf {\bibinfo {volume} {74}},\ \bibinfo
  {pages} {063806} (\bibinfo {year} {2006})}\BibitemShut {NoStop}%
\bibitem [{\citenamefont {Sun}\ \emph {et~al.}(2008)\citenamefont {Sun},
  \citenamefont {Böttger}, \citenamefont {Thiel},\ and\ \citenamefont
  {Cone}}]{sun_magnetic_2008}%
  \BibitemOpen
  \bibfield  {author} {\bibinfo {author} {\bibfnamefont {Y.}~\bibnamefont
  {Sun}}, \bibinfo {author} {\bibfnamefont {T.}~\bibnamefont {B{\"o}ttger}},
  \bibinfo {author} {\bibfnamefont {C.~W.}\ \bibnamefont {Thiel}},\ and\
  \bibinfo {author} {\bibfnamefont {R.~L.}\ \bibnamefont {Cone}},\ }\bibfield
  {title} {\bibinfo {title} {Magnetic {$g$} tensors for the {$^4$I$_{15/2}$}
  and {$^4$I$_{13/2}$} states of {Er$^{3+}$:Y$_2$SiO$_5$}},\ }\href
  {https://doi.org/10.1103/PhysRevB.77.085124} {\bibfield  {journal} {\bibinfo
  {journal} {Physical Review B}\ }\textbf {\bibinfo {volume} {77}},\ \bibinfo
  {pages} {085124} (\bibinfo {year} {2008})}\BibitemShut {NoStop}%
\bibitem [{\citenamefont {Rueda}\ \emph {et~al.}(2016)\citenamefont {Rueda},
  \citenamefont {Sedlmeir}, \citenamefont {Collodo}, \citenamefont {Vogl},
  \citenamefont {Stiller}, \citenamefont {Schunk}, \citenamefont {Strekalov},
  \citenamefont {Marquardt}, \citenamefont {Fink}, \citenamefont {Painter},
  \citenamefont {Leuchs},\ and\ \citenamefont
  {Schwefel}}]{rueda_efficient_2016}%
  \BibitemOpen
  \bibfield  {author} {\bibinfo {author} {\bibfnamefont {A.}~\bibnamefont
  {Rueda}}, \bibinfo {author} {\bibfnamefont {F.}~\bibnamefont {Sedlmeir}},
  \bibinfo {author} {\bibfnamefont {M.~C.}\ \bibnamefont {Collodo}}, \bibinfo
  {author} {\bibfnamefont {U.}~\bibnamefont {Vogl}}, \bibinfo {author}
  {\bibfnamefont {B.}~\bibnamefont {Stiller}}, \bibinfo {author} {\bibfnamefont
  {G.}~\bibnamefont {Schunk}}, \bibinfo {author} {\bibfnamefont {D.~V.}\
  \bibnamefont {Strekalov}}, \bibinfo {author} {\bibfnamefont {C.}~\bibnamefont
  {Marquardt}}, \bibinfo {author} {\bibfnamefont {J.~M.}\ \bibnamefont {Fink}},
  \bibinfo {author} {\bibfnamefont {O.}~\bibnamefont {Painter}}, \bibinfo
  {author} {\bibfnamefont {G.}~\bibnamefont {Leuchs}},\ and\ \bibinfo {author}
  {\bibfnamefont {H.~G.~L.}\ \bibnamefont {Schwefel}},\ }\bibfield  {title}
  {\bibinfo {title} {Efficient microwave to optical photon conversion: An
  electro-optical realization},\ }\href
  {https://doi.org/10.1364/OPTICA.3.000597} {\bibfield  {journal} {\bibinfo
  {journal} {Optica}\ }\textbf {\bibinfo {volume} {3}},\ \bibinfo {pages} {597}
  (\bibinfo {year} {2016})}\BibitemShut {NoStop}%
\bibitem [{\citenamefont {Pound}(1946)}]{pound_electronic_1946}%
  \BibitemOpen
  \bibfield  {author} {\bibinfo {author} {\bibfnamefont {R.~V.}\ \bibnamefont
  {Pound}},\ }\bibfield  {title} {\bibinfo {title} {Electronic {{Frequency
  Stabilization}} of {{Microwave Oscillators}}},\ }\href
  {https://doi.org/10.1063/1.1770414} {\bibfield  {journal} {\bibinfo
  {journal} {Review of Scientific Instruments}\ }\textbf {\bibinfo {volume}
  {17}},\ \bibinfo {pages} {490} (\bibinfo {year} {1946})}\BibitemShut
  {NoStop}%
\bibitem [{\citenamefont {Neuhaus}\ \emph {et~al.}(2017)\citenamefont
  {Neuhaus}, \citenamefont {Metzdorff}, \citenamefont {Chua}, \citenamefont
  {Jacqmin}, \citenamefont {Briant}, \citenamefont {Heidmann}, \citenamefont
  {Cohadon},\ and\ \citenamefont {Del{\'e}glise}}]{neuhaus_pyrpl_2017}%
  \BibitemOpen
  \bibfield  {author} {\bibinfo {author} {\bibfnamefont {L.}~\bibnamefont
  {Neuhaus}}, \bibinfo {author} {\bibfnamefont {R.}~\bibnamefont {Metzdorff}},
  \bibinfo {author} {\bibfnamefont {S.}~\bibnamefont {Chua}}, \bibinfo {author}
  {\bibfnamefont {T.}~\bibnamefont {Jacqmin}}, \bibinfo {author} {\bibfnamefont
  {T.}~\bibnamefont {Briant}}, \bibinfo {author} {\bibfnamefont
  {A.}~\bibnamefont {Heidmann}}, \bibinfo {author} {\bibfnamefont {P.-F.}\
  \bibnamefont {Cohadon}},\ and\ \bibinfo {author} {\bibfnamefont
  {S.}~\bibnamefont {Del{\'e}glise}},\ }\bibfield  {title} {\bibinfo {title}
  {{{PyRPL}} ({{Python Red Pitaya Lockbox}}) - an open-source software package
  for {{FPGA-controlled}} quantum optics experiments},\ }in\ \href
  {https://doi.org/10.1109/CLEOE-EQEC.2017.8087380} {\emph {\bibinfo
  {booktitle} {2017 {{European Conference}} on {{Lasers}} and
  {{Electro-Optics}} and {{European Quantum Electronics Conference}}}}}\
  (\bibinfo {year} {2017})\ p.\ \bibinfo {pages} {EA\_P\_8}\BibitemShut
  {NoStop}%
\bibitem [{\citenamefont {Yu}\ \emph {et~al.}(2016)\citenamefont {Yu},
  \citenamefont {Janousek}, \citenamefont {Sheridan}, \citenamefont {McAuslan},
  \citenamefont {{Rubinsztein-Dunlop}}, \citenamefont {Lam}, \citenamefont
  {Zhang},\ and\ \citenamefont {Bowen}}]{yu_optomechanical_2016}%
  \BibitemOpen
  \bibfield  {author} {\bibinfo {author} {\bibfnamefont {C.}~\bibnamefont
  {Yu}}, \bibinfo {author} {\bibfnamefont {J.}~\bibnamefont {Janousek}},
  \bibinfo {author} {\bibfnamefont {E.}~\bibnamefont {Sheridan}}, \bibinfo
  {author} {\bibfnamefont {D.~L.}\ \bibnamefont {McAuslan}}, \bibinfo {author}
  {\bibfnamefont {H.}~\bibnamefont {{Rubinsztein-Dunlop}}}, \bibinfo {author}
  {\bibfnamefont {P.~K.}\ \bibnamefont {Lam}}, \bibinfo {author} {\bibfnamefont
  {Y.}~\bibnamefont {Zhang}},\ and\ \bibinfo {author} {\bibfnamefont {W.~P.}\
  \bibnamefont {Bowen}},\ }\bibfield  {title} {\bibinfo {title} {Optomechanical
  {{Magnetometry}} with a {{Macroscopic Resonator}}},\ }\href
  {https://doi.org/10.1103/PhysRevApplied.5.044007} {\bibfield  {journal}
  {\bibinfo  {journal} {Physical Review Applied}\ }\textbf {\bibinfo {volume}
  {5}},\ \bibinfo {pages} {044007} (\bibinfo {year} {2016})}\BibitemShut
  {NoStop}%
\bibitem [{\citenamefont {Li}\ \emph {et~al.}(2012)\citenamefont {Li},
  \citenamefont {Lee}, \citenamefont {Yang},\ and\ \citenamefont
  {Vahala}}]{li_sideband_2012}%
  \BibitemOpen
  \bibfield  {author} {\bibinfo {author} {\bibfnamefont {J.}~\bibnamefont
  {Li}}, \bibinfo {author} {\bibfnamefont {H.}~\bibnamefont {Lee}}, \bibinfo
  {author} {\bibfnamefont {K.~Y.}\ \bibnamefont {Yang}},\ and\ \bibinfo
  {author} {\bibfnamefont {K.~J.}\ \bibnamefont {Vahala}},\ }\bibfield  {title}
  {\bibinfo {title} {Sideband spectroscopy and dispersion measurement in
  microcavities},\ }\href {https://doi.org/10.1364/OE.20.026337} {\bibfield
  {journal} {\bibinfo  {journal} {Optics Express}\ }\textbf {\bibinfo {volume}
  {20}},\ \bibinfo {pages} {26337} (\bibinfo {year} {2012})}\BibitemShut
  {NoStop}%
\bibitem [{\citenamefont {Mlynek}\ \emph {et~al.}(1983)\citenamefont {Mlynek},
  \citenamefont {Wong}, \citenamefont {DeVoe}, \citenamefont {Kintzer},\ and\
  \citenamefont {Brewer}}]{mlynek_raman_1983}%
  \BibitemOpen
  \bibfield  {author} {\bibinfo {author} {\bibfnamefont {J.}~\bibnamefont
  {Mlynek}}, \bibinfo {author} {\bibfnamefont {N.~C.}\ \bibnamefont {Wong}},
  \bibinfo {author} {\bibfnamefont {R.~G.}\ \bibnamefont {DeVoe}}, \bibinfo
  {author} {\bibfnamefont {E.~S.}\ \bibnamefont {Kintzer}},\ and\ \bibinfo
  {author} {\bibfnamefont {R.~G.}\ \bibnamefont {Brewer}},\ }\bibfield  {title}
  {\bibinfo {title} {Raman {{Heterodyne Detection}} of {{Nuclear Magnetic
  Resonance}}},\ }\href {https://doi.org/10.1103/PhysRevLett.50.993} {\bibfield
   {journal} {\bibinfo  {journal} {Phys. Rev. Lett.}\ }\textbf {\bibinfo
  {volume} {50}},\ \bibinfo {pages} {993} (\bibinfo {year} {1983})}\BibitemShut
  {NoStop}%
\bibitem [{\citenamefont {{Fernandez-Gonzalvo}}\ \emph
  {et~al.}(2015)\citenamefont {{Fernandez-Gonzalvo}}, \citenamefont {Chen},
  \citenamefont {Yin}, \citenamefont {Rogge},\ and\ \citenamefont
  {Longdell}}]{fernandez-gonzalvo_coherent_2015}%
  \BibitemOpen
  \bibfield  {author} {\bibinfo {author} {\bibfnamefont {X.}~\bibnamefont
  {{Fernandez-Gonzalvo}}}, \bibinfo {author} {\bibfnamefont {Y.-H.}\
  \bibnamefont {Chen}}, \bibinfo {author} {\bibfnamefont {C.}~\bibnamefont
  {Yin}}, \bibinfo {author} {\bibfnamefont {S.}~\bibnamefont {Rogge}},\ and\
  \bibinfo {author} {\bibfnamefont {J.~J.}\ \bibnamefont {Longdell}},\
  }\bibfield  {title} {\bibinfo {title} {Coherent frequency up-conversion of
  microwaves to the optical telecommunications band in an {{Er}}:{{YSO}}
  crystal},\ }\href {https://doi.org/10.1103/PhysRevA.92.062313} {\bibfield
  {journal} {\bibinfo  {journal} {Physical Review A}\ }\textbf {\bibinfo
  {volume} {92}},\ \bibinfo {pages} {062313} (\bibinfo {year}
  {2015})}\BibitemShut {NoStop}%
\bibitem [{\citenamefont {King}\ \emph {et~al.}(2021)\citenamefont {King},
  \citenamefont {Barnett}, \citenamefont {Bartholomew}, \citenamefont
  {Faraon},\ and\ \citenamefont {Longdell}}]{king_probing_2021}%
  \BibitemOpen
  \bibfield  {author} {\bibinfo {author} {\bibfnamefont {G.~G.~G.}\
  \bibnamefont {King}}, \bibinfo {author} {\bibfnamefont {P.~S.}\ \bibnamefont
  {Barnett}}, \bibinfo {author} {\bibfnamefont {J.~G.}\ \bibnamefont
  {Bartholomew}}, \bibinfo {author} {\bibfnamefont {A.}~\bibnamefont
  {Faraon}},\ and\ \bibinfo {author} {\bibfnamefont {J.~J.}\ \bibnamefont
  {Longdell}},\ }\bibfield  {title} {\bibinfo {title} {Probing strong coupling
  between a microwave cavity and a spin ensemble with {{Raman}} heterodyne
  spectroscopy},\ }\href {https://doi.org/10.1103/PhysRevB.103.214305}
  {\bibfield  {journal} {\bibinfo  {journal} {Physical Review B}\ }\textbf
  {\bibinfo {volume} {103}},\ \bibinfo {pages} {214305} (\bibinfo {year}
  {2021})}\BibitemShut {NoStop}%
\end{thebibliography}

%apsrev4-2.bst 2019-01-14 (MD) hand-edited version of apsrev4-1.bst
%Control: key (0)
%Control: author (8) initials jnrlst
%Control: editor formatted (1) identically to author
%Control: production of article title (0) allowed
%Control: page (0) single
%Control: year (1) truncated
%Control: production of eprint (0) enabled
%

\end{document}